\documentclass[11pt,letterpaper]{article}
\usepackage{times}
\usepackage[onehalfspacing]{setspace}
\usepackage{natbib}\bibpunct{(}{)}{,}{}{}{,}
\usepackage{amsmath,amsfonts,amsthm}
\usepackage{comment}
\usepackage{tabularx}
\usepackage{multirow}
\usepackage{booktabs}
\usepackage{subcaption}
\usepackage{graphicx}
\usepackage[colorlinks,linkcolor=blue,citecolor=black,urlcolor=black]{hyperref}
\usepackage[title,titletoc]{appendix}
\usepackage{enumitem}
\usepackage{subcaption}  % For subfigures
\usepackage{lscape}      % For landscape orientation if needed
\usepackage[noabbrev,capitalize]{cleveref}

\newcommand{\figpath}{fig/}

\begin{document}

\title{\textbf{Exports, Labor Markets, and the Environment: Evidence from Brazil}\thanks{This paper is a product of the Poverty and Equity Practice at the World Bank Group. We gratefully acknowledge the financial support of Climate Trust Fund “The impact of the green transition on employment and livelihoods of poor and vulnerable households”. We thank colleagues from Brazil's Ministry of Development, Industry, Commerce and Services and the Secretariat of Monitoring and Evaluation of Public Policies for the insightful discussions, as well as Nancy Lozano for insightful comments. The views expressed herein are those of the authors and do not necessarily reflect the views of the World Bank.}}
\author{\large%
\setcounter{footnote}{0}%
Carlos G\'{o}es \\[-3pt] \textit{\small IFC, World Bank Group}%
\and\setcounter{footnote}{1}%
Otavio Concei\c{c}\~ao%
\\[-3pt] \textit{\small World Bank}%
\and%
Gabriel Lara Ibarra%
\\[-3pt] \textit{\small World Bank}
\and%
Gladys Lopez-Acevedo%
 \\[-3pt] \textit{\small World Bank}%
}
\maketitle

\vspace{-0.7cm}
\noindent\begin{abstract}\begin{singlespace}
\noindent What is the environmental impact of exports? Focusing on 2000–20, this paper combines customs, administrative, and census microdata to estimate employment elasticities with respect to exports. The findings show that municipalities that faced increased exports experienced faster growth in formal employment. The elasticities were 0.25 on impact, peaked at 0.4, and remained positive and significant even 10 years after the shock, pointing to a long and protracted labor market adjustment. In the long run, informal employment responds negatively to export shocks. Using a granular taxonomy for economic activities based on their environmental impact, the paper documents that environmentally risky activities have a larger share of employment than environmentally sustainable ones, and that the relationship between these activities and exports is nuanced. Over the short run, environmentally risky employment responds more strongly to exports relative to environmentally sustainable employment. However, over the long run, this pattern reverses, as the impact of exports on environmentally sustainable employment is more persistent.\\[4pt]\textbf{Keywords}: exports shocks, labor markets, formal employment, wages, Brazil.
\\[4pt] \textbf{JEL Classification}: D3, F16, J16, O19
\end{singlespace}\end{abstract}

\newpage
\section{Introduction}\label{sec:intro}
What is the environmental impact of exports? The answer is not straightforward. For instance, international trade can incentivize firms to adopt cleaner technologies and more efficient production methods, either through competitive pressures or environmental regulation spillovers. At the same time, reallocation of production due to cost-based comparative advantage from places with relatively “greener” input matrices to others with relatively “dirtier” ones may harm the environment, if those factors are not internalized in market prices. 
Even in theory, depending on critical modeling assumptions such as the relationship between carbon intensity across sectors and firm productivity, there is no unique answer for whether emissions increase or decrease after trade liberalization \citep{wanner_new_2024}. The outcome will depend on scale, composition, and technique effects induced by trade \citep{cherniwchan_international_2022}.

In this paper, we approach this question through a labor market lens and estimate how exports impact employment in sectors with different environmental characteristics. Focusing on Brazil, a country that experienced a large increase in exports over the 2000s, we combine customs, administrative, and census microdata with a granular taxonomy of activities and estimate employment elasticities with respect to exports for different subgroups. There is a widespread perception that the success of the exporting sector has been associated with the destruction of Brazilian ecosystems  \citep{martinelli_agriculture_2010}. However, the evidence regarding the direct impact of exports on environmental outcomes is mixed.\footnote{\cite{felbermayr_trade_2024} present a review of this relationship. \cite{dornelas_china_2019} finds that the effect of increased exports due to Chinese demand on deforestation of the Amazon rainforest and Cerrado savannah was on average insignificant. Similarly, \cite{carreira_deforestation_2024} show that the same shock induced replacing cropland for pasture rather than deforestation.} Exploring the cross-group and cross-industry heterogeneities sheds light on the nexus between the Brazilian export boom and its environmental impact.

We start by estimating the formal employment elasticity to exports over time. To answer the question of interest, we construct a panel of municipalities with information on formal employment and exports ranging from 2000 to 2020. To account for potential endogeneity of exports, we construct a shift-share instrument leveraging pre-treatment local labor market composition across different industries and the growth of global exports in each sector. Prior to the arrival of an export shock, there are no differential employment trends between municipalities more exposed to relative to those less exposed to the shock. When the shock occurs, more exposed municipalities experience faster growth in formal employment. We calculate elasticities to be 0.25 on impact and to peak at 0.4 three years after the shock. Consistent with a positive demand shock, wage elasticities are also positive, hovering about 0.2 over the medium term.

To shed light on the relationship between exports and the environment, we use a very granular taxonomy of activities to classify employment in each municipality according to environmental characteristics. We label economic activities subject to regulatory environmental impact licensing as environmentally risky activities and those that significantly contribute to environmental objectives, such as climate change mitigation or adaptation, as environmentally sustainable ones.

We document that environmentally risky activities have a larger share of employment than those that are environmentally sustainable. In 2010, the median municipality had 20 percent of its workforce allocated to environmentally sustainable activities and 50 percent of its workforce allocated to environmentally risky activities. In estimating the employment responses of each of those employment groups to exports, we find a nuanced picture. Environmentally risky activities in municipalities more exposed to export shocks have relatively higher growth over the short run but that effect fades away over time. Conversely, the response of environmentally sustainable employment is more long lasting, with elasticities above 0.17 even 10 years after the initial shock.

One of the shortcomings of administrative data that we use to estimate the employment elasticity is that it only covers formal employment, excluding about half of the Brazilian labor force. To overcome this limitation, we use two waves of decennial census data to separately estimate elasticities of overall, formal and informal employment. Ten years after the initial shock, while municipalities more exposed to export shocks see relatively higher growth in formal employment, they experience a relative decrease in informal employment. Our results are consistent with the interpretation that, in developing countries, formal and informal labor employment are substitutes and informal labor markets act as a mechanism to accommodate shocks. Finally, we show that over the long term, environmentally risky employment elasticities are positive for formal employment but negative for informal employment. Conversely, while formal employment in environmentally sustainable activities also has positive long-run elasticities, informal environmentally sustainable employment is unaffected in the long run.

This paper contributes to the literature in at least three ways. 

First, we document a protracted and long-lasting labor market adjustment after positive export shocks. While there is a large literature documenting slow labor market adjustment after trade shocks, most of it focuses on import competition or input cost reduction shocks.\footnote{For instance, \cite{topalova_factor_2010} measured changes in tariff rates by weighting industry-level shifts based on worker distribution in each Indian district, applying the Bartik-type approach to explore the relationship between globalization and local labor market outcomes. She found that districts with a larger share of import-competing sectors, and those facing greater tariff reductions, were more exposed to trade liberalization shocks, with tariff reductions viewed as exogenous, being set by central government trade agreements. Famously, \cite{autor_china_2013} show that US commuting zones that faced larger import competition from China experienced lower employment growth relative to those less impacted by import competition. Additionally, labor market adjustment has been long-lasting.}
Our work is related to \cite{costa_winners_2016}, who find that local labor markets more affected by rising Chinese commodity demand between 2000 and 2010 experienced faster wage growth. Similarly, \cite{brummund_labor_2018} examined Brazil’s trade relationship with China, finding that export exposure reduced transitions from the traded sector to non-employment and increased shifts from non-employment to the non-traded sector, particularly in manufacturing.

Second, we relate trade shocks to labor market outcomes in countries with a large informal labor market and confirm that formal and informal employment tend to act as substitutes in Brazil after large trade shocks. Our work is related to \cite{dix-carneiro_trade_2017}, who found that in the long run, foreign competition had no effect on unemployment, instead significantly increasing relative informal employment in regions more exposed to it. Likewise, \cite{ponczek_enforcement_2022} confirmed that the medium-term unemployment effects of liberalization-induced foreign competition were greater in regions with stricter labor regulations, which hampered labor market flexibility and hindered shifts in employment. As a result, the informal sector emerged as a critical buffer for displaced workers.

Third, we estimate the potential impact of exports on the environment from a local labor market perspective. Our work contributes to the literature that estimates non-economic outcomes of trade policy variation. \cite{bombardini_trade_2020} show that localities more exposed to increased exports in China faced a relatively higher increase of pollution-induced mortality through a sectoral composition channel. \cite{dornelas_china_2019} concludes that the rise in exports driven by Chinese demand had on average, an insignificant impact on deforestation. Likewise, \cite{carreira_deforestation_2024} demonstrate that this export shock did not lead to deforestation. 

The paper is structured as follows. Section 2 provides an overview of Brazil's trade and labor market trends, including novel facts about how the environmental characteristics of the labor force changed over time. Section 3 outlines the methodology and data construction. Section 4 discusses the main findings from the shift-share analysis, focusing on the relationship between export growth, labor market outcomes, and the environment. Section 5 concludes with a summary of the key insights.

\section{Trade, environment, and the labor market in Brazil: Recent trends}\label{sec:context}
\begin{figure}[htp]
    \centering
    \includegraphics[width=0.7\linewidth]{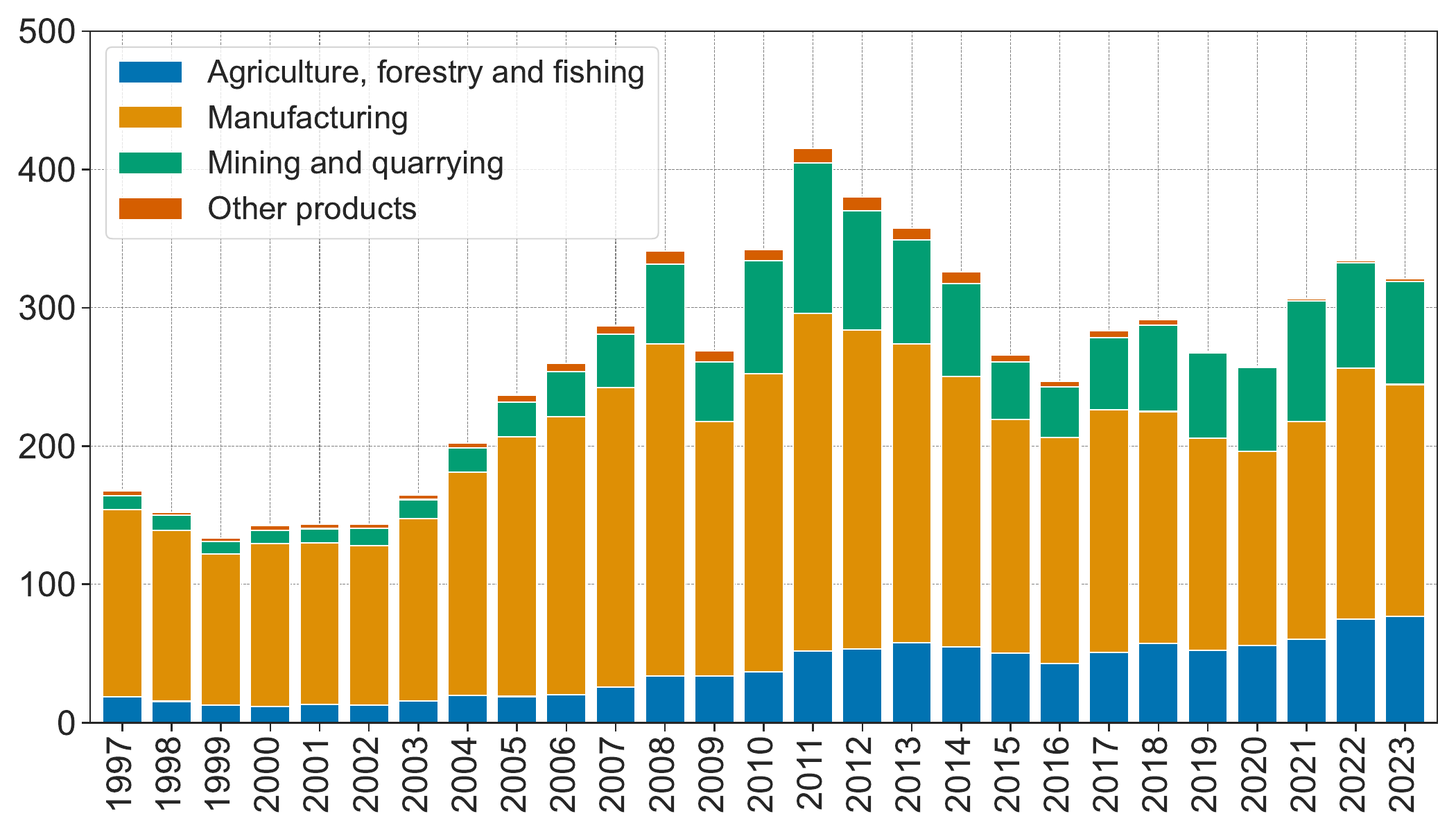}
    \caption{Evolution of Brazilian exports, separately by sector \\ {\footnotesize Source: elaborated by the authors using data from the Brazilian Ministry of Industry and Commerce (MDIC), the Brazilian NSO (IBGE) and the Federal Reserve Economic Data (FRED). Notes: Values are denominated in billions of U.S. dollars at 2022 prices. }}
    
    \label{fig:exports-time-series}
\end{figure}

Over the last 25 years, Brazil’s exports of goods increased by about three times in real terms. Exports were close to USD 170 billion (at 2022 prices) in 1997, peaked at USD 400 billion in 2010 and declined to slightly more than USD 300 billion in 2023. Figure \ref{fig:exports-time-series} depicts the historic evolution considering the 1997-2023 period with data separately for some categories (Agriculture, Forestry and Fishing, Manufacturing, and Mining and quarrying). There has been, thus, an increase in exports in real terms comparing 1997 with 2023, but with an important reduction from 2010 to 2023 of about 25 percent. Overall, we note that the trend of the Brazilian export cycle in this period is a combination of a continuous expansion of the agricultural sector, with a large cycle of oil and a volatile manufacturing sector.

\begin{figure}[ht!]
    \centering
    \begin{minipage}{0.45\textwidth}
        \centering
        \includegraphics[width=\textwidth]{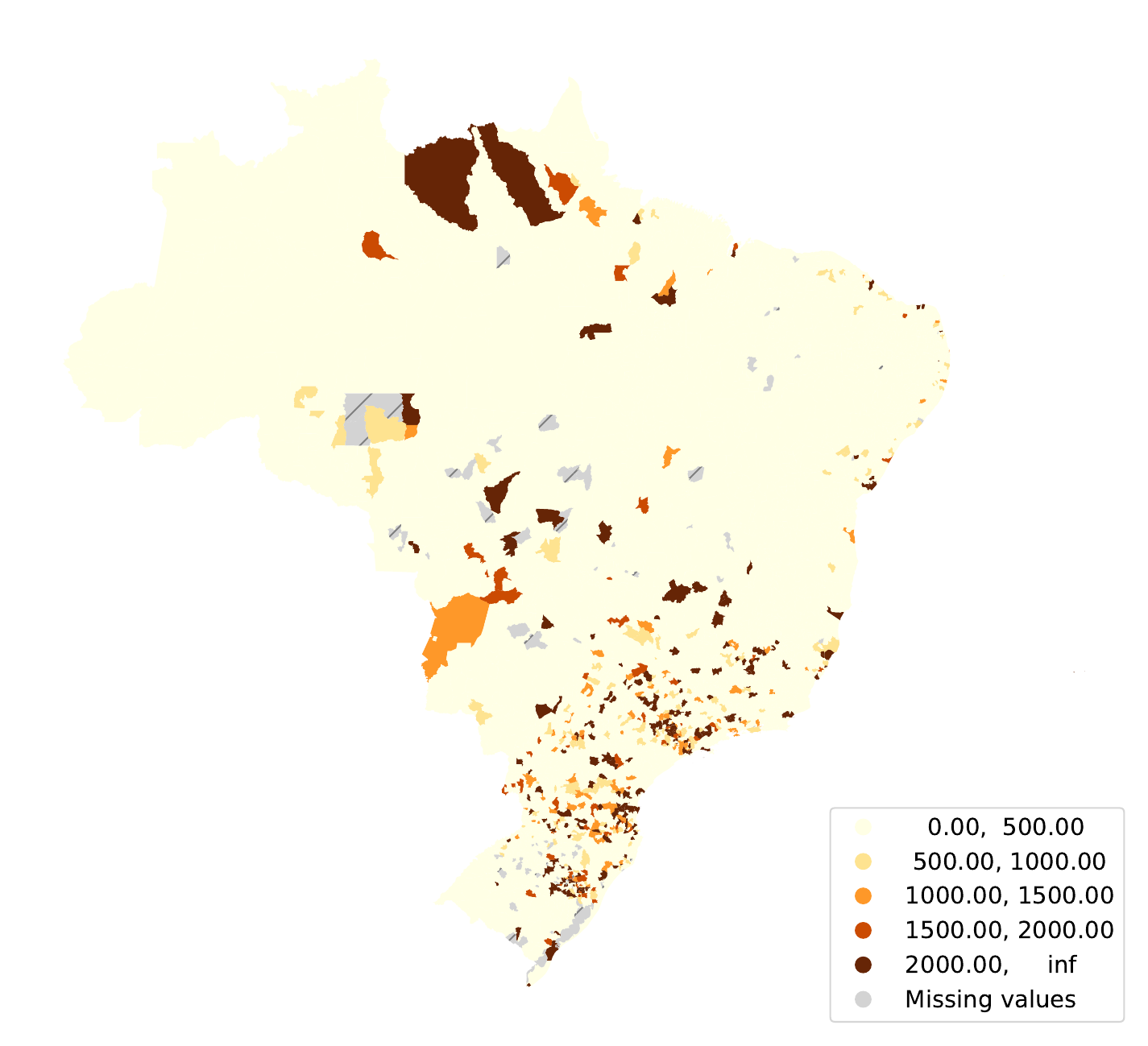} % Third figure
        \\
        2000
    \end{minipage}%
    \hspace{0.04\textwidth}
    \begin{minipage}{0.45\textwidth}
        \centering
        \includegraphics[width=\textwidth]{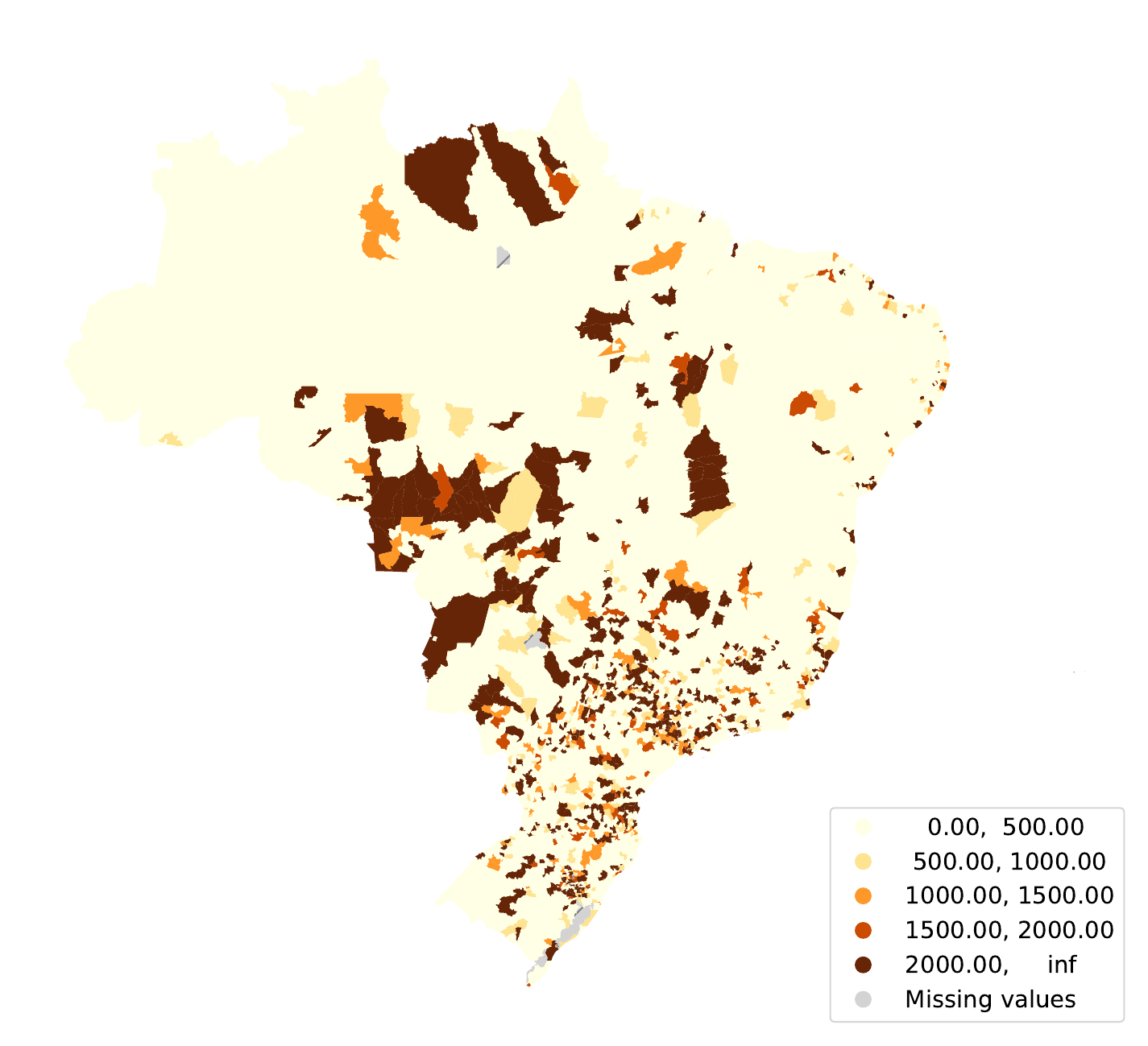} % Fourth figure
        \\
        2010
    \end{minipage}

    \caption{Spatial distribution of real exports per person at the municipality level (constant 2022 USD). \\
    {\footnotesize Source: elaborated by the authors using data from the Brazilian Ministry of Industry and Commerce (MDIC), the Brazilian NSO (IBGE) and the Federal Reserve Economic Data (FRED). Notes: values in the label (to the right of each map) are denominated in U.S. dollars at 2022 prices.}}
    \label{fig:map-exports}
\end{figure}

Figure \ref{fig:map-exports} presents the distribution of exports per person in U.S. dollars in 2020 free-on-board prices in Brazilian municipalities both in 2000 and 2010 – before and after the exports boom. The vast majority of Brazilian municipality do not export. In 2000, about 28 percent of the municipalities had positive exports. This share grew to 33 percent in 2010 and more than 40 percent in 2020. By 2010, about 8 percent of municipalities exported at least USD 2,000 per person, a near three-fold percentage point increase since 2000. The spatial distribution of exports also changed substantially over time. In 2000, exports were concentrated in the traditional manufacturing hubs of the Southeast and South. By 2010, while they were still important, the Midwest and parts of the Northeast had a larger imprint.

\begin{figure}[htp]
    \centering
    \includegraphics[width=0.7\linewidth]{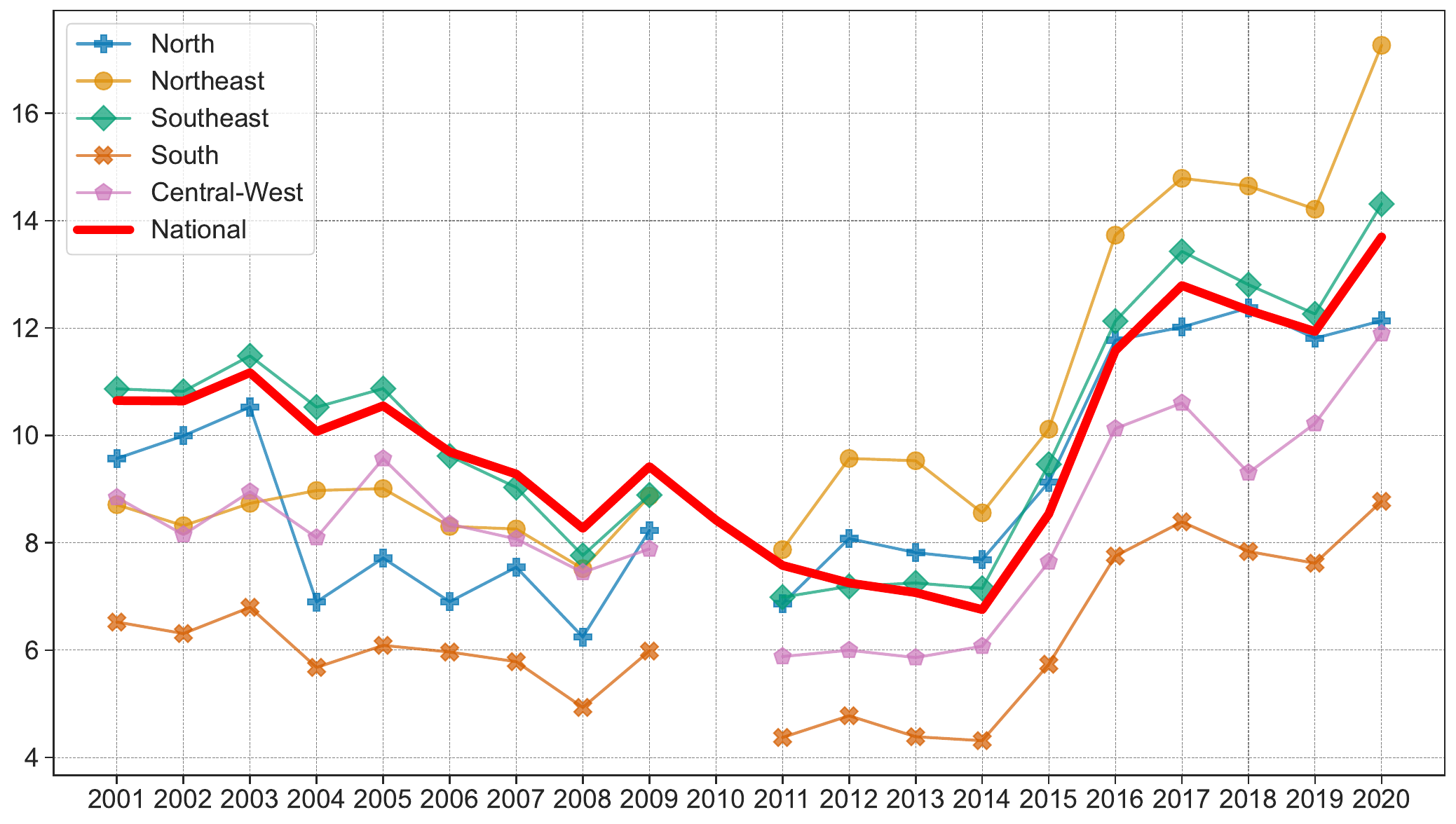}
    \caption{Unemployment rate (percent) by region in Brazil, 2001-2020 \\ {\footnotesize Source: elaborated by the authors using data from IBGE. Notes: The data for 2001-2009 and 2011 comes from PNAD, while that for 2012-2020 comes from PNAD-C. The national household survey (PNAD) was not fielded in 2010 because of the 2010 Population Census. We decided not to include the 2010 data from the census because of comparability issues involving census and the national household surveys.  }}
    
    \label{fig:unemployment}
\end{figure}

The Brazilian labor market has also undergone significant transformations over the past two decades, characterized by regional, gender, and ethnic disparities. The economic boom of the early 2000s led to substantial job creation, particularly in the Southeast and South regions, where industrial and service sectors thrived. This period also witnessed a notable reduction in poverty rates and increased female labor force participation. However, the benefits of this growth were unevenly distributed, with the Northeast and North regions lagging in terms of job creation and income levels. The economic downturn that began in 2015 has exacerbated existing regional disparities, enlarging gaps in unemployment rates across areas (Figure \ref{fig:unemployment}).

\begin{figure}[ht!]
    \centering

    % Panel A (two figures side by side in one row)
    \textbf{Panel A: Environmentally risky activities (share of total employment)} \\[1em] % Header for Panel A
    \begin{minipage}{0.45\textwidth}
        \centering
        \includegraphics[width=\textwidth]{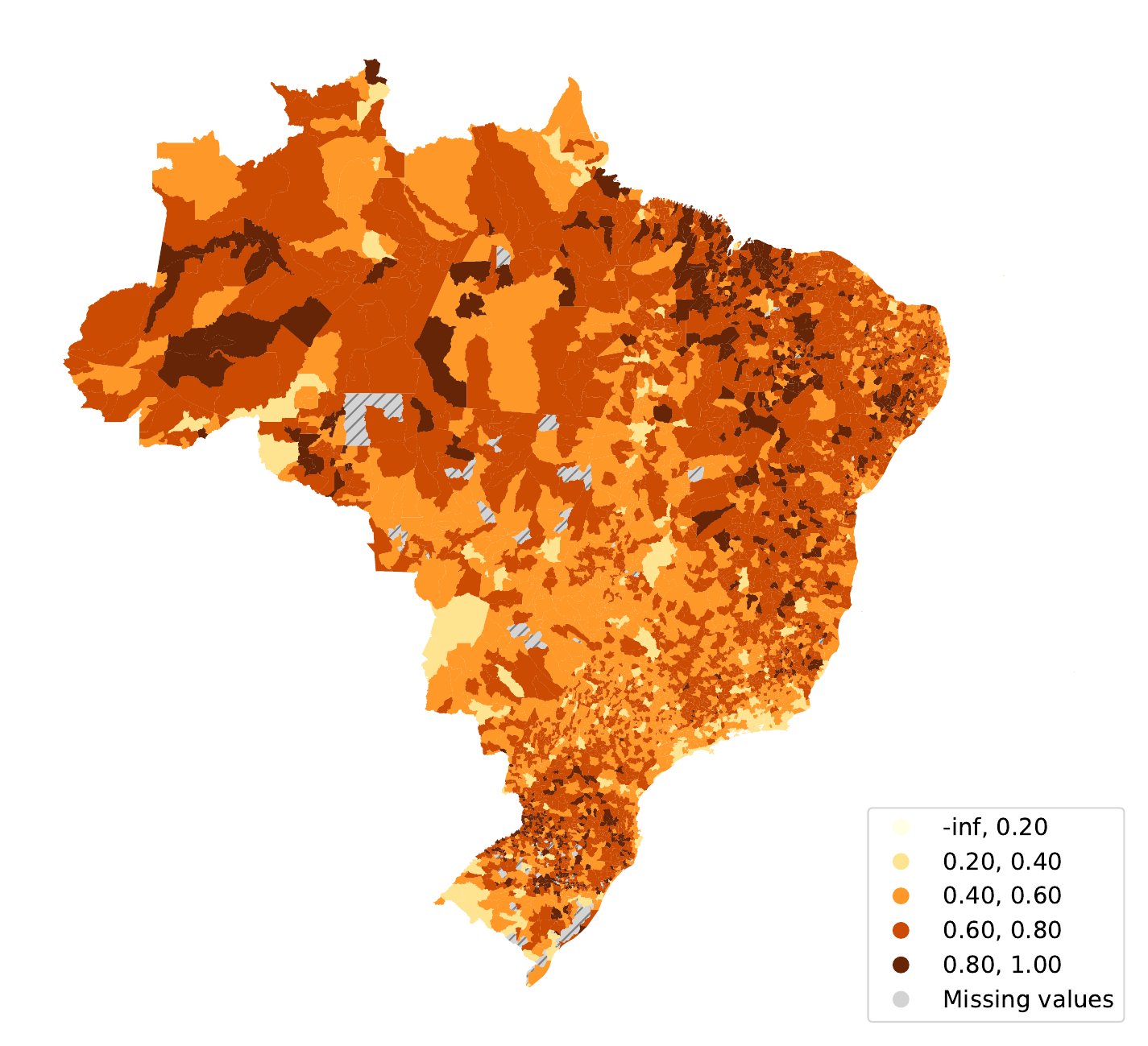} % First figure 
        \\
        2000
    \end{minipage}%
    \hspace{0.04\textwidth}
    \begin{minipage}{0.45\textwidth}
        \centering
        \includegraphics[width=\textwidth]{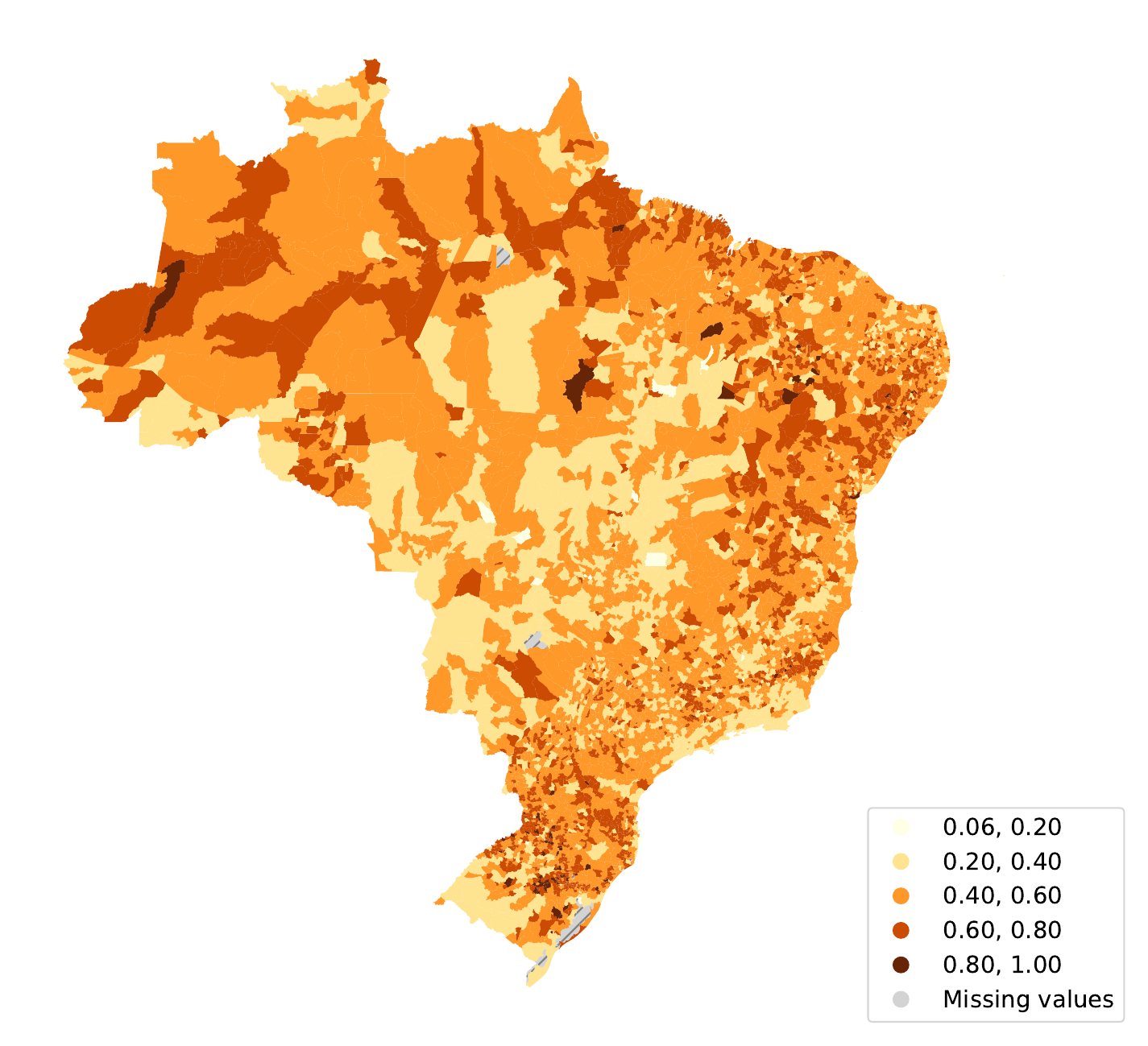} % Second figure
        \\
        2010
    \end{minipage}
    
    \vspace{1em} % Vertical space between the panels

    % Panel B (two figures side by side in one row)
    \textbf{Panel B: Environmentally sustainable activities (share of total employment)} \\[1em] % Header for Panel B
    \begin{minipage}{0.45\textwidth}
        \centering
        \includegraphics[width=\textwidth]{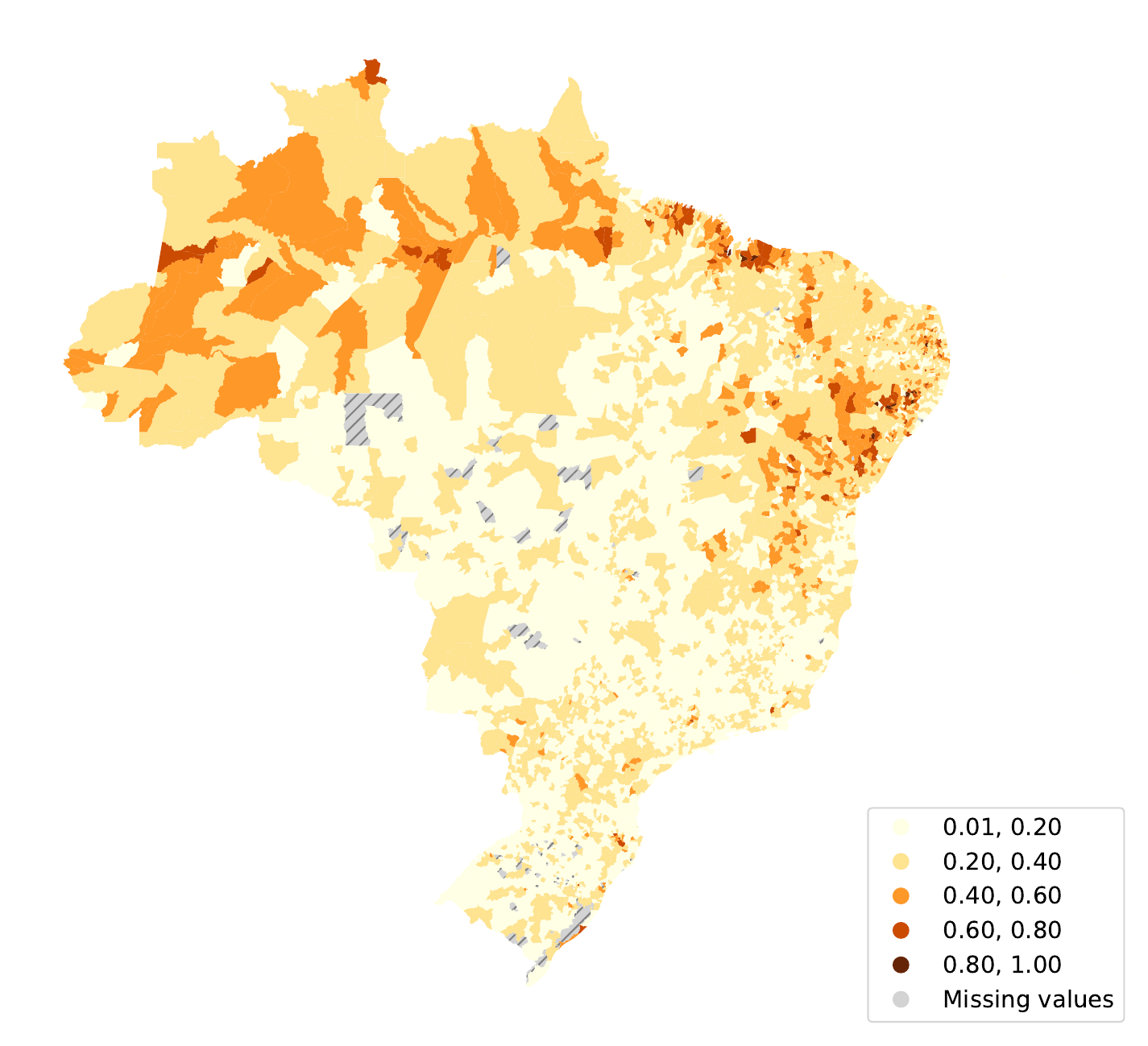} % Third figure
        \\
        2000
    \end{minipage}%
    \hspace{0.04\textwidth}
    \begin{minipage}{0.45\textwidth}
        \centering
        \includegraphics[width=\textwidth]{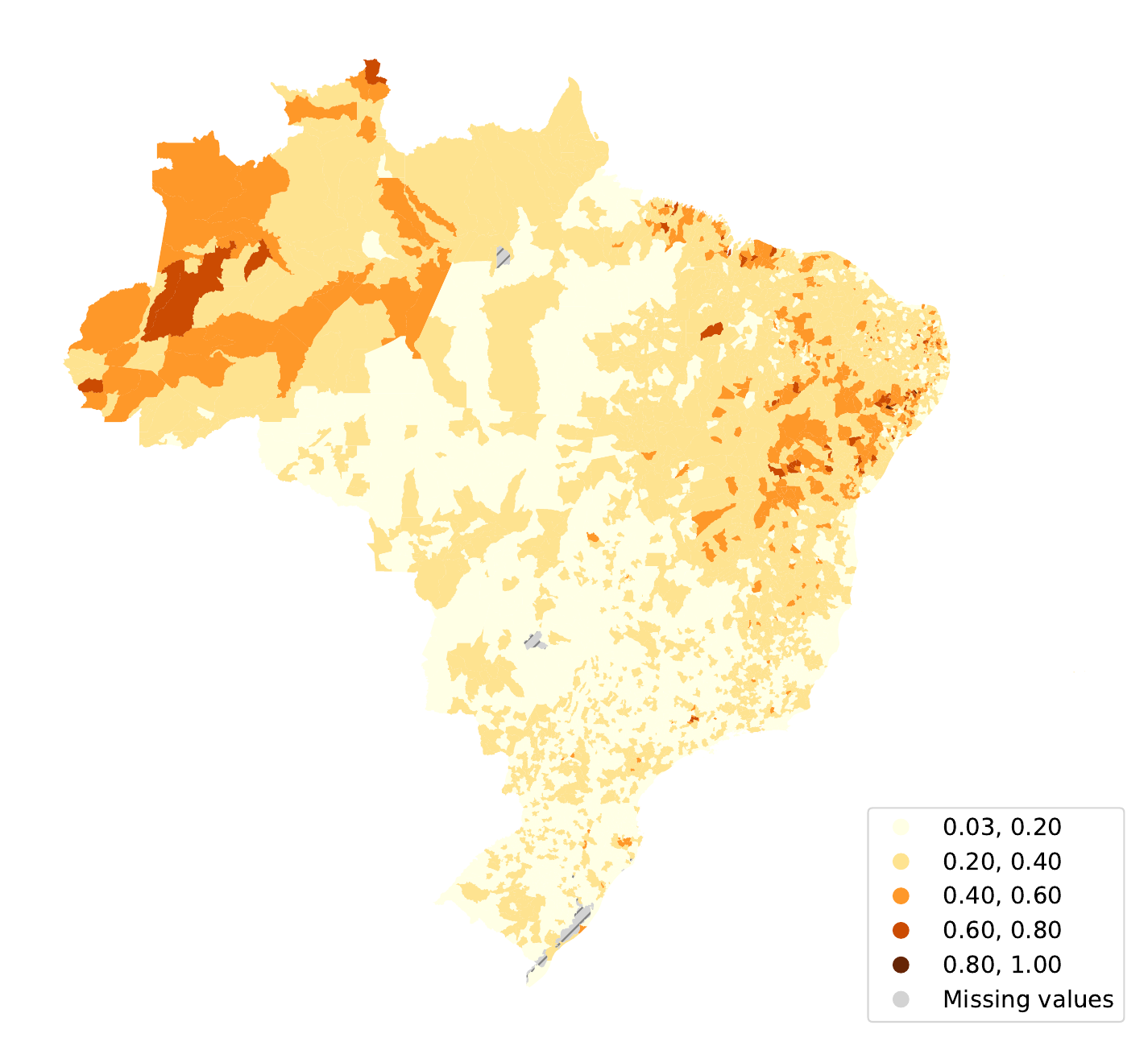} % Fourth figure
        \\
        2010
    \end{minipage}

    \caption{Spatial distribution of environmental patterns of the labor force. \\
    {\footnotesize Source: elaborated by the authors using data from Febraban and the Brazilian NSO (IBGE).}}
    \label{fig:map-taxonomy}
\end{figure}

While the Brazilian economy has shown signs of recovery between 2017 and 2019, the labor market continued to face challenges. The service sector, dominated by female employment, has been a major driver of job creation, but the quality of these jobs often remains precarious. The gap in unemployment rates between white and Afro Brazilian workers persists, reflecting historical inequalities (World Bank, 2022), and informality remains a major issue, with informal employment accounting for 62.4 percent in 2023 with very limited variation compared to 2013. 
By combining a granular taxonomy of activities with labor market data, we document two novel facts about how the environmental patterns of the labor force evolve across time and space in Brazil. Below, we plot such evolution using decennial Census data, which covers both the formal and informal labor markets. First, workers in environmentally risky activities (i.e., those for which an environmental permit is required) are common and widespread in all regions (Panel A of Figure \ref{fig:map-taxonomy}). Importantly, the share of workers allocated to those activities decreased over time: the average across municipalities was 65 percent of total employment in 2000 and 50 percent in 2010. The national aggregate share of employment in environmentally risky activities decreased from 44 to 32 percent in the same period.

Second, workers who perform environmentally sustainable activities (i.e., those related to renewable energy transition; carbon pricing; biodiversity protection; circular economy; zero pollution; sustainable agriculture and food systems; improving forests and land use; and improving water and marine protection) have a relatively lower share of total employment and are concentrated in specific parts of the country (Panel B of Figure \ref{fig:map-taxonomy}). The municipalities around the Amazon and São Francisco rivers stand out as having a higher share. The distribution across municipalities did not change much over time, with an average increase of 1 percentage point between 2000 (19 percent) and 2010 (20 percent). The national aggregate share of employment in environmentally sustainable activities remained stable at 21 percent in the same period.

The shares of environmentally risky and environmentally sustainable employment are negatively correlated across municipalities ($\rho= -0.18$) and across national aggregate sectors ($\rho= -0.1$). This suggests that the taxonomy is indeed capturing different dimensions. 

\newpage

\section{Data and methodology}\label{sec:methods}
\paragraph{Motivation} Many classes of structural models of international trade result in a similar expenditure equation with a gravity structure \citep{arkolakis_new_2012}. Following GDP accounting, they result in an equation that states that total national income of region $r$ equals total expenditure on goods produced in region $r$. Under usual assumptions, when there are many industries $k$, the equation will take the following form:

\begin{equation*}
    w_r L_r \propto \sum_k Y_{r,k} = \sum_k \sum_d X_{rd,k} = \sum_k \sum_d \frac{T_{r,k} (\tau_{rd,k}c_{r,k})^{-\theta_k}}{\Phi_{d,k}} X_{d,k} = \sum_k \sum_d \lambda_{rd,k} X_{d,k}
\end{equation*}

where $w_r L_r$ is total labor income and $Y_r$ is total income in region $r$; total income can be further decomposed into expenditure on goods produced by industry $k$ of region $r$ $Y_{r,k}$ as well as total expenditure in industry $k$ at destination region $d$ in goods produced in $r$ $X_{rd,k}$; $T_{r,k}$ is a term that captures the productivity of sector $k$ in region $r$; $c_{r,k}$ is the unit input cost of industry $k$ in region $r$; $\tau_{rd,k}\ge 1$ captures the transport cost of shipping a good of industry $k$ between $r$ and $d$, and is meant to capture both geographical distances and tariffs; $\Phi_{d,k} = \sum_{s'} T_{s',k} (\tau_{s'd,k}c_{s',k})^{-\theta_k}$ captures the average cost of goods in industry $k$ at the destination region $d$; $\lambda_{rd,k}$ is the expenditure trade share on goods coming from region $r$ out of total expenditure in destination $d$ spent on sector $k$; and $-\theta_k$ is the trade elasticity, i.e., the responsiveness of trade flows with respect to changes in trade costs. Define $\hat{x} = dx / x$. Then, totally differentiating the equation above results in:

\begin{equation*}
    \hat{L}_r + \hat{w}_r\propto \sum_{k} \mu_{r,k} \hat{Y}_{r,k} 
\end{equation*}

\noindent where $ \mu_{r,k} \equiv Y_{r,k} / Y_r$ is the initial weight of industry $k$. Using the definition of $Y_{r,k}$ and following through with total differentiation, we can write:

\begin{equation*}
    \hat{L}_r + \hat{w}_r\propto \sum_k \mu_{r,k} \sum_{d} \lambda_{rd,k} \left[ \left( \hat{T}_{r,k} - \sum_{s'} \lambda_{s'd,k} \hat{T}_{s',k} \right) - \theta_{k} \left( \hat{c}_{r,k} + \hat{\tau}_{rd,k} - \sum_{s'} \lambda_{s'd,k} \left( \hat{c}_{s',k} + \hat{\tau}_{s'd,k} \right) \right) + \hat{X}_{d,k} \right]
\end{equation*}

The first term within the brackets captures relative changes in supply shocks. This channel has been the focus of much of the literature of labor market adjustment to trade, notably by 
\cite{autor_china_2013} and \cite{pierce_surprisingly_2016}. Both papers focused on the effects of the ``China trade shock'' across local labor markets, leveraging differential exposure to Chinese import penetration due to plausibly exogenous increases in Chinese productivity.

The second term captures relative changes in trade and production costs. Most of the literature that uses Brazilian administrative data emphasizes this mechanism. For instance, \cite{kovak_regional_2013}, \cite{dix-carneiro_trade_2014}, and \cite{felix_trade_2022} all concentrate on a vector of industry-specific tariff reductions interacted with local-labor market specific exposure weights to trace out labor market adjustment.

We focus instead on the last term within the brackets. This is the impact of changes in global demand in industry $k$ due to changes in preferences, sector prices or aggregate income. If region $r$ is small relative to the world, changes to foreign demand are a plausibly exogenous source of variation of exports in $r$.

\paragraph{Empirical strategy}  Using customs administrative data, we observe exports flows $\tilde{X}_{r,t}$ in each local labor market $r$ in Brazil for many periods $t$. Using the equations in the motivation subsection as a reference, we map total expenditure to exports as: $\tilde{X}_{r,t} = \sum_k (Y_{r,k} - \sum_{d \in \mathcal{B}} X_{rd,k})$, where $\mathcal{B} := \{r:r \text{ is a region of Brazil}\}$. The growth in exports is $\Delta \tilde{x}_{r,t}$, where $\tilde{x}_{r,t}$ is the log of exports of region $r$ in period $t$.

If exports were as good as random, we would be able to recover the treatment effect of an increase of expenditures on $r$ by regressing some outcome of interest on exports. However, there are many reasons to believe that exposure to exports can be endogenous. Shocks to domestic local technology $\hat{T}_{r,k}$ and costs $\hat{c}_{r,k} + \hat{\tau}_{rd,k}$ drive exports and can be naturally correlated with unobserved local labor market characteristics. Therefore, one needs to use some plausibly exogenous instrument uncorrelated to domestic demand and supply shocks to consistently recover this effect.

Once again referring to the stylized structural model, global demand shifters are transmitted to domestic income as: $\sum_k \mu_{r,k} \sum_{d\notin \mathcal{B}} \lambda_{rd,k} \hat{X}_{d,k}$. Like \cite{aghion_heterogeneous_2018}, we use world export flows in each industry as a demand shock. As the industry weights in a local labor market, we follow \cite{autor_china_2013} and \cite{dix-carneiro_trade_2014}, among others, and use local labor market shares.\footnote{Assuming free labor mobility across sectors, then $L_{r,g} = Y_{r,g} /w_r  $ for every $g$. Summing over $g$, this implies that $L_r = \sum_{g}Y_{r,g}/w_r$, and, therefore: $L_{r,g} / L_r = Y_{r,g} / \sum_{g}Y_{r,g}$, which maps from industry shares to labor shares under these assumptions.} Our measure of exposure to global demand shocks in region $r$ is, then:

\begin{equation}\label{eq: baseline-instrument}
    \Delta \bar{x}_{r,t} \equiv \sum_k \frac{L_{r,k,t-1}}{L_{r,t-1}} \Delta \tilde{x}^w_{k,t}
\end{equation}

\noindent where $L_{r,k,t-1}$ denotes total formal employment in industry $k$ at region $r$ at period $t-1$; $L_{r,t-1} = \sum_k L_{r,k,t-1}$; and $\Delta \tilde{x}^w_{k,t}$ is the growth of world exports (other than Brazil’s) in industry $k$. In the robustness section, we consider an instrument that leverages changes in dollar income in destination countries.

We are concerned about trade adjustment dynamics. Our approach is to estimate a sequence of local projection regressions as in \cite{jorda_estimation_2005}. Given some outcome of interest $\{o_{r,t}\}$ and a vector of control variables $\{\textbf{Z}_{r,t}'\}$, estimation takes the form of two-stage least squares, with the first stage being:

\begin{equation*}
    \Delta \tilde{x}_{r,t} = \alpha + \beta \Delta \bar{x}_{r,t} + \textbf{Z}_{r,t-1}' \Gamma + \varepsilon_{r,t}
\end{equation*}

\noindent and the second stage:

\begin{equation}\label{eq:main-model}
    o_{r,t+h} - o_{r,t-1} = \alpha_h + \beta_h \Delta \hat{x}_{r,t} + \textbf{Z}_{r,t-1}' \Gamma_h + \varepsilon_{r,t+h} \qquad \text{for } h \in \{ \cdots, -1, 0, 1,  \cdots \}
\end{equation}

\noindent where $\Delta \hat{x}_{r,t}$ are the predicted values of the first stage regression. 

In this sequence of regressions, for each $t$, the right-hand variables are fixed at the time of the shock. The dependent variable, which changes for each horizon, denotes the cumulative change of the outcome variable since the base period. The coefficients $\beta_h$ form a cumulative impulse response function, denoting the dynamic average treatment effect of the outcome variable. The vector of controls $\textbf{Z}_{r,t-1}' = [\Delta o_{r,pre},w_{r,2000},s_{r,2000}]$ will account for potential differential pre-trends $\Delta o_{r,pre} = o_{r,t-2} - o_{r,t-5}$, convergence (initial sample-period real wages, $w_{r,2000}$), and export share composition in labor market industries (we control for the initial share of industries that do not export in each region $s_{r,2000}=\sum_{k \text{ does not export}} L_{k,r,2000} / L_{r,2000}$).

In reduced form, we account for the labor market impacts of exports induced either directly due to foreign demand shocks or indirectly through how changes in demand affect other equilibrium objects.

\begin{figure}[htp]
    \centering
    \includegraphics[width=0.75\linewidth]{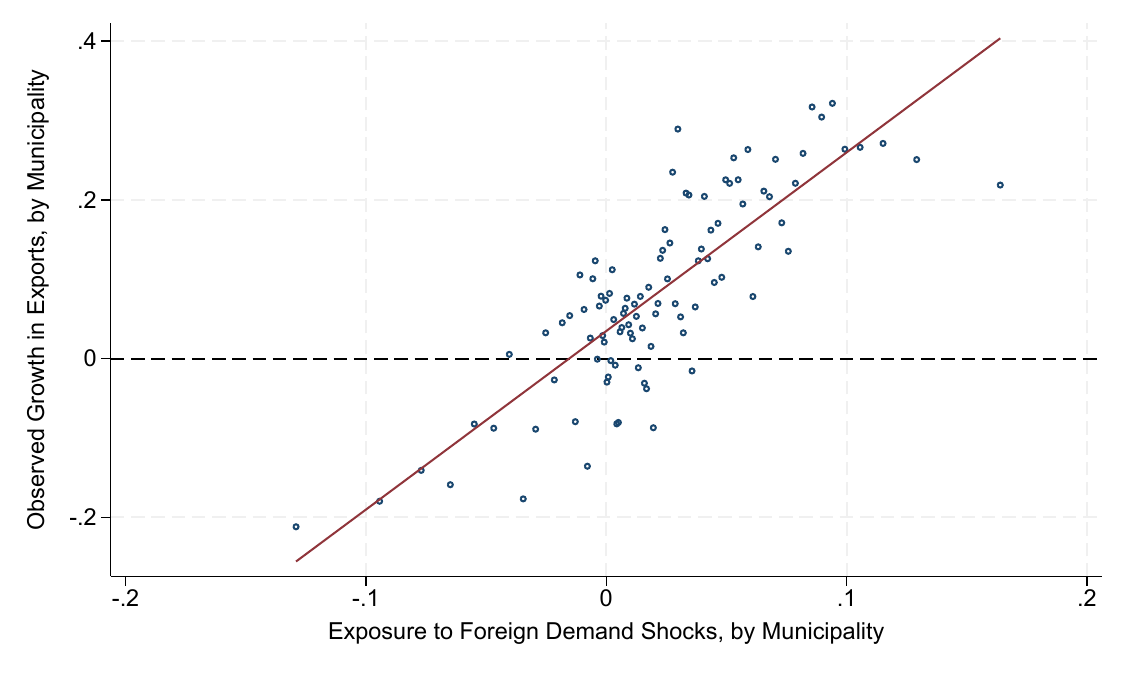}
    \caption{Relevance of the instrumental variable \\ {\footnotesize Notes: this binscatter reproduces the slope of regressing the observed growth in exports on the instrument. The underlying regression has N=34,670, Beta = 2.25 and t-stat=16.76. }}
    \label{fig:first-stage}
\end{figure}

For the instrument to be relevant, regional exposure of industry-specific global demand shocks needs to be strongly correlated with observed growth in exports. Reassuringly, the F-statistic of the proposed instrument in the first-stage regression is greater than 280, which is remarkably high and suggests a nonnegligible correlation. To further inspect the relevance of the instrument, Figure \ref{fig:first-stage} depicts a binscatter where the instrument is presented in the horizontal axis, while the endogenous variable is shown in the vertical axis using municipality-level data. As can been seen, there is an unequivocally strong and positive relationship between those variables and a relatively low dispersion of observations around the fitted line.

For validity, the requirement is that our choice of instrument only affects the outcome variables through its impacts on the instrumented variable – i.e., the exclusion restriction. Estimation of $\beta_h$ is consistent if $\mathbb{E}[\Delta \bar{x}_{r,t} \cdot \varepsilon_{r,h} \mid \textbf{Z}_{r,t-1}, L_{r,k,t-1}]=0$ for every $d$ and $r$ pair at every horizon $h$; that is, if changes in foreign demand are uncorrelated with the distribution of unobserved factors that drive changes in local labor markets. In the language of the shift-share literature, we take shares as given and derive identification from plausibly exogenous shifters \citep{borusyak_quasi-experimental_2020}\footnote{One alternative approach would be to every share as individual instruments, each operating under the same difference-in-differences logic. See \cite{borusyak_practical_2025, goldsmith-pinkham_bartik_2020}}.

\paragraph{The estimator} Our empirical approach leverages regional differences in exposure to exports. However, such cross-sectional variation strategies only capture the effect of increased export exposure relative to other regions. In fact, this estimator is akin to a differences-in-differences estimator, with an additional term that accounts for spillover effects between regions, which violate the Stable Unit Treatment Value Assumption (SUTVA). Following \cite{chodorow-reich_geographic_2019}, define the relative treatment effect at horizon $h$ as:

\begin{equation*}
    \beta_h \equiv \frac{ O_{r,t+h}( [\textbf{W}_{t-1} + i_r\Delta_h]', W_{agg,t-1} ) -  O_{r,t+h}( \textbf{W}_{t-1}', W_{agg,t-1} ) }{\Delta_h}
\end{equation*}

\noindent 
where $O_{r,t+h}( \textbf{W}_{t-1}', W_{agg,t-1})$ is the potential outcome function of region $r$ at period $t+h$, for some arbitrary treatment arguments; $\textbf{W}_t= (W_{1,t},\cdots,W_{N,t} )'$ is the vector of region-specific treatments; $i_r$ is a unit vector with one in row $r$ and zeros elsewhere; and $W_{agg,t-1}$ is some aggregate treatment that affects the regions uniformly, such as exchange rate interventions or monetary policy following a commodity boom. Note that since the potential outcome function depends on the whole vector of individual treatments, it can potentially account for potential spillovers. However, in using the unit vector $i_r$, we are defining the relative treatment effect $\beta_h$ is the marginal impact of treatment dosage $\Delta_h$ on region $r$, holding all other individual and aggregate treatments constant.

When a single unit is treated, the empirical object estimated in regional analysis is:

\begin{equation*}
    \hat{\beta}_h \equiv \frac{ [o_{r,t+h} - o_{r,t-1}] - (N-1)^{-1} \sum_{i \neq r}[o_{i,t+h} - o_{i,t-1}]}{\Delta_h}
\end{equation*}

For $\hat{\beta}_h \to \beta_h$, when a single region is treated, sufficient conditions for consistency is that either each region has mass zero ($N \to \infty$) or that potential outcomes for non-treated units do not change when region $r$ is treated ($\partial o_{i,t-1} / \partial W_{r,t-1} = 0$ $\forall i \neq r$) – i.e., SUTVA to hold. In short, regional regressions can be interpreted as difference-in-differences (DiD) design.

In our setting, units are treated through a weighted average of aggregate industry-level shocks with regional exposure defined by pre-existing labor market shares. Recent developments in the shift-share literature explain that such parallel with DiDs persists when generalized to multiple regions being treated\footnote{For a comprehensive review of this literature, see \citep{borusyak_practical_2025}}.

As in DiDs, one should interpret the estimates as \textit{relative} rather than \textit{aggregate} effects. 
Above, since $W_{agg,t-1}$ is assumed to uniformly impact all regions, it has no impact on $\beta_h$ \textemdash for this reason, regional regressions suffer from a ``missing intercept problem'' \citep{wolf_missing_2021} with respect to general equilibrium effects and can only capture relative effects. One cannot extrapolate from the cross-sectional to the aggregate effect without imposing additional assumptions. Likewise, as in DiDs, identification relies on treatment timing and differential exposure. As such, one should perform balance tests to ensure that the instrument is uncorrelated with pre-treatment covariates and assess the plausibility of endogeneity. 

\paragraph{Labor data} Our primary data source to analyze labor market dynamics in Brazil is a matched employer-employee dataset known as RAIS (Relação Anual de Informações Sociais). RAIS is an annual census of formal workers administered by the Brazilian Ministry of Labor, containing detailed information about nearly the universe of formal employees in the country. Employers are required to submit information about their employees to RAIS every year and face penalties for non-compliance with submission deadlines, ensuring high accuracy of reported information. We observe employees\footnote{Natural Persons Registry, \textit{Cadastro de Pessoas Físicas}.} and firm\footnote{National Registry of Legal Entities, \textit{Cadastro Nacional de Pessoas Jurídicas}.} unique tax identifiers and utilize municipality and industry codes of firms and wages and demographic characteristics of workers to aggregate data at the region-industry-year level. 

Those in the informal labor market are not included in RAIS. Hence, to complement the analysis, we use censuses data for the waves of 2000 and 2010. Census questionnaires aim to cover the entire population, and we can use their data to examine both formal and informal labor markets at the municipality-sector level.\footnote{While census data is available from as early as 1960, our focus is on the most recent period of export expansion in Brazil, so we restrict our analysis to the 2000 and 2010 censuses. We do not include the 1991 census because the sectoral classification used in RAIS data was established after the 1991 census, and no official concordance exists between the census 1991 classification and the RAIS classification. Creating an ad-hoc concordance would require numerous assumptions, potentially introducing additional noise into the analysis.} The formal labor market is defined based on whether a worker has a formal job contract (i.e., \textit{carteira assinada}) or contributes to social security. Census data allow us to use the same levels of spatial aggregation as in RAIS. Annual labor market surveys are not representative at geographic levels of aggregation finer than 26 Brazilian states in additional to the Federal District. We instead use data at after aggregating it into 5,571 municipalities.

\paragraph{Trade data} Brazilian trade data comes from customs records. The Ministry of Industry and Commerce of Brazil (MDIC) publishes customs records at the municipality level. Location of imports and exports are recorded based on the address reported by the importing/exporting firm. Flows are classified for each International Standard Industrial Classification (ISIC) revision 3 3-digit industry code.\footnote{Municipal exports are published at the Harmonized System (HS) 4-digit product level and 3-digit ISIC industry level. We also have product level data at the HS 8-digit level (Nomenclatura Comum do Mercosul - NCM) for flows at the national and state level, but those are not available at the municipality level due to fiscal secrecy laws.}  We observe flows quantities and Free on Board (FOB) nominal dollar values and use total FOB dollar values as a metric of municipal exports.

For constructing the baseline instrument, we use data from the UN COMTRADE database regarding global exports, excluding Brazilian exports, mapped to ISIC 3-digit industry code.\footnote{UN COMTRADE data are published at the HS 6-digit product level. We use the concordances from the United Nations Statistics Division to harmonize the HS vintages and map them to ISIC rev. 3 3-digit industry codes.} Finally, we map Brazilian employment data (based on 5-digit industry classification) to trade data by aggregating employment indicators (i.e. level, wages, etc.) in each municipality to the ISIC 3-digit industry code.\footnote{We used the Brazilian national statistical office (IBGE) concordance between the national classification of economic activities (CNAE) at a 3-digit industry level and ISIC rev. 3 3-digit industry code.} There are 111 different sectors at the 3-digit industry classification.

\paragraph{Environmental taxonomy} To define environmentally risky and environmentally sustainable activities, we use the taxonomy developed by the Brazilian Federation of Banks (FEBRABAN) that is tailored to Brazil and very granular. The details of this taxonomy are explained below. Appendix \ref{sec:app-febraban} provides additional information about how taxonomy was mapped to labor data.  

First, we define a set of environmentally risky activities. An economic activity is defined at the 5-digit industry level. We classify activity $a$ of a 3-digit industry $k$ (i.e. sector) as risky if they are subject to environmental impact analysis prior to receiving a permit to operate. This risk classification, defined by Resolution  237/1997 from the Brazilian National Council on Environment (CONAMA), precedes our period of analysis and is plausibly exogenous to the (future) foreign demand shocks we consider. The set of environmentally risky activities for each sector $k$ is, then:

\begin{equation*}
    \mathcal{R}(k) := \{ a : \text{ activity } a \text{ of sector } k \text{ is subject to environmental impact analysis}\}  
\end{equation*}

\begin{figure}[ht!]
    \centering

    % Panel A (two figures side by side in one row)
    \textbf{Panel A: Change over time in environmental characteristics of workers} \\[1em] % Header for Panel A
    \begin{minipage}{0.45\textwidth}
        \centering
        \includegraphics[width=\textwidth]{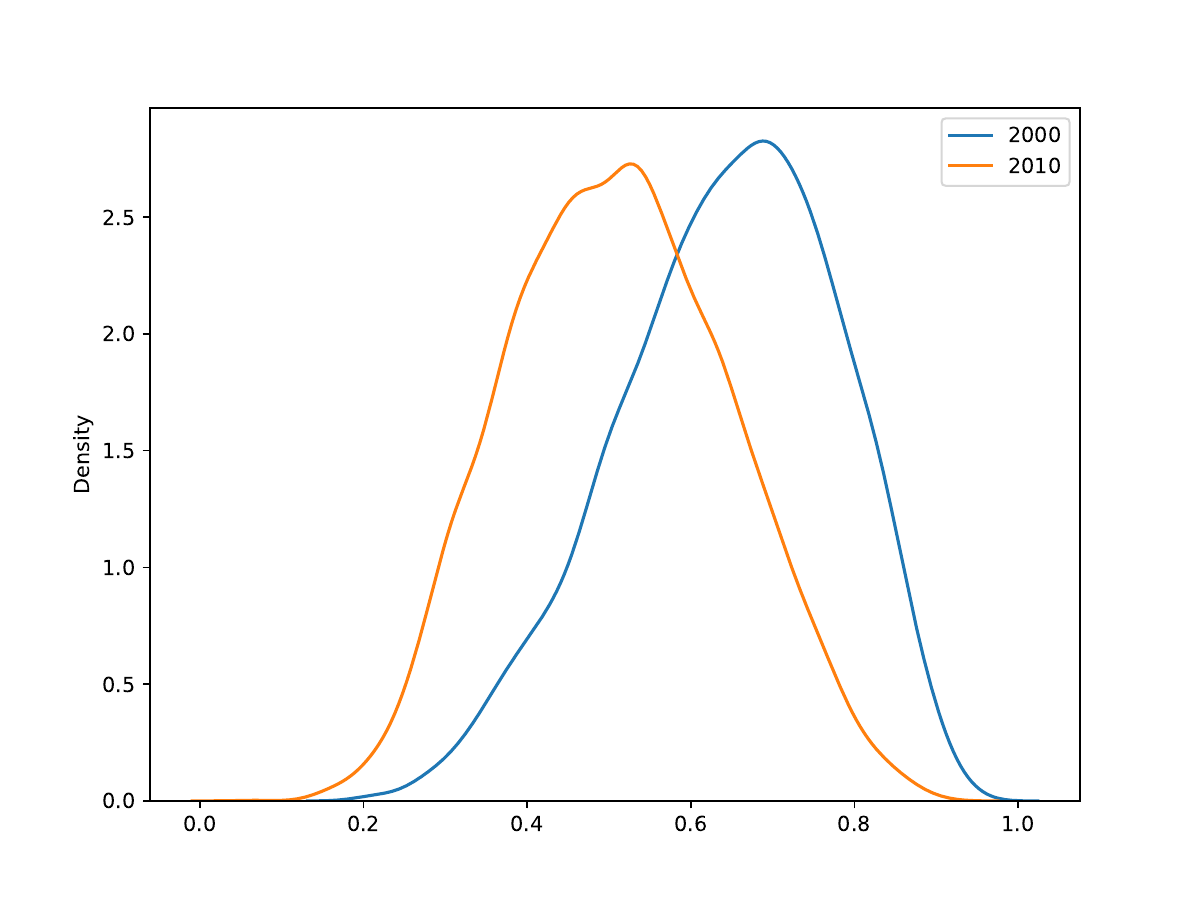} % First figure 
        \\
        Environmentally risky activities, 2000-10 (all workers)
    \end{minipage}%
    \hspace{0.04\textwidth}
    \begin{minipage}{0.45\textwidth}
        \centering
        \includegraphics[width=\textwidth]{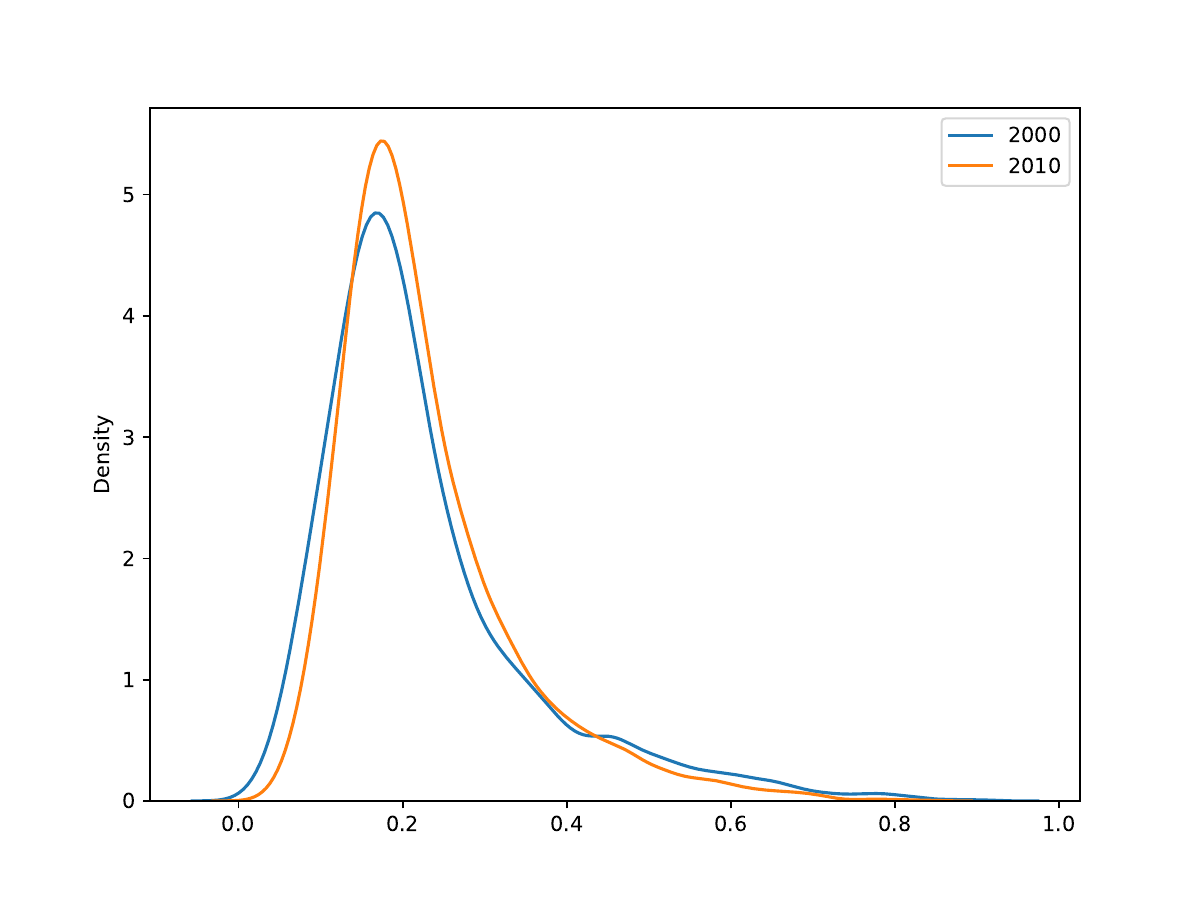} % Second figure
        \\
        Environmentally sustainable activities, 2000-10  (all workers)
    \end{minipage}
    
    \vspace{1em} % Vertical space between the panels

    % Panel B (two figures side by side in one row)
    \textbf{Panel B: Differences in environmental characteristics of workers by formalization} \\[1em] % Header for Panel B
    \begin{minipage}{0.45\textwidth}
        \centering
        \includegraphics[width=\textwidth]{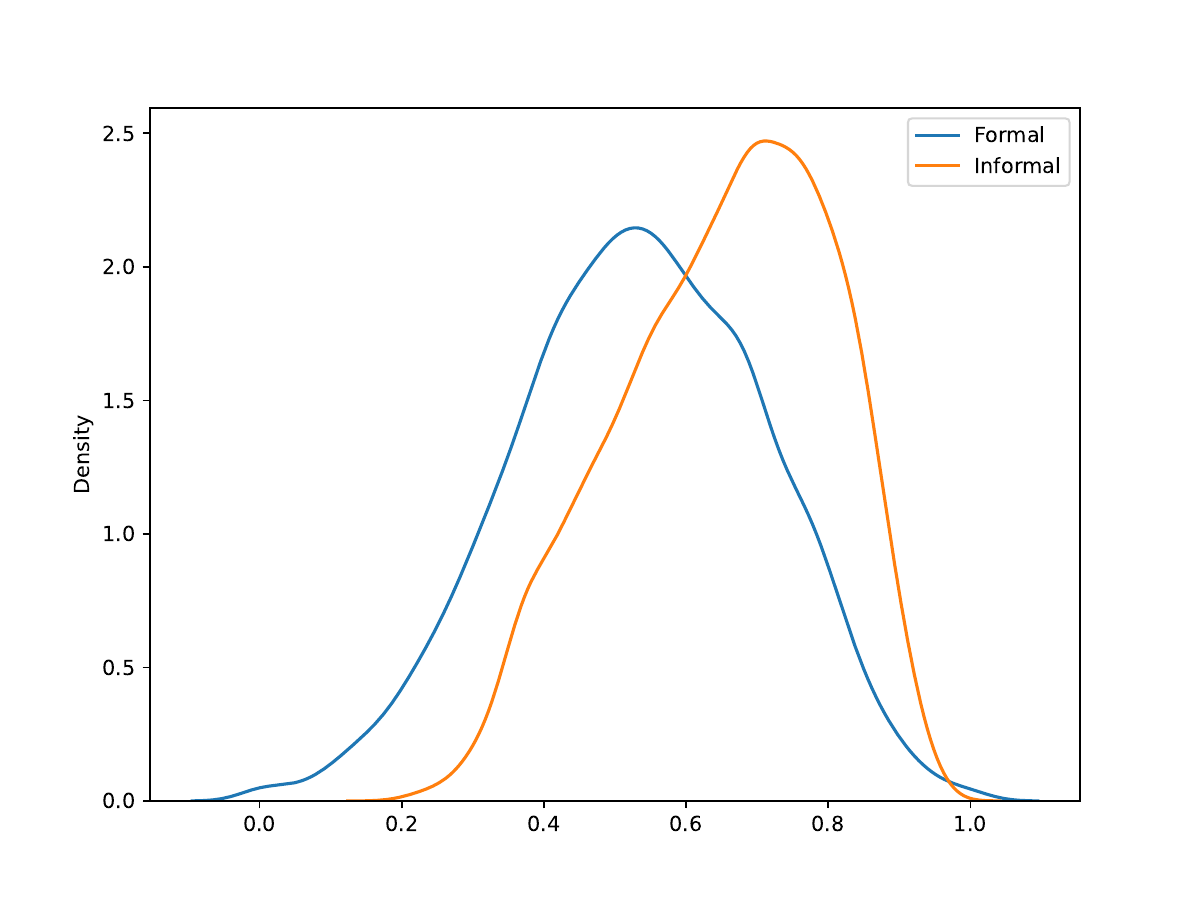} % Third figure
        \\
        Environmentally risky activities, 2000 (by sector)
    \end{minipage}%
    \hspace{0.04\textwidth}
    \begin{minipage}{0.45\textwidth}
        \centering
        \includegraphics[width=\textwidth]{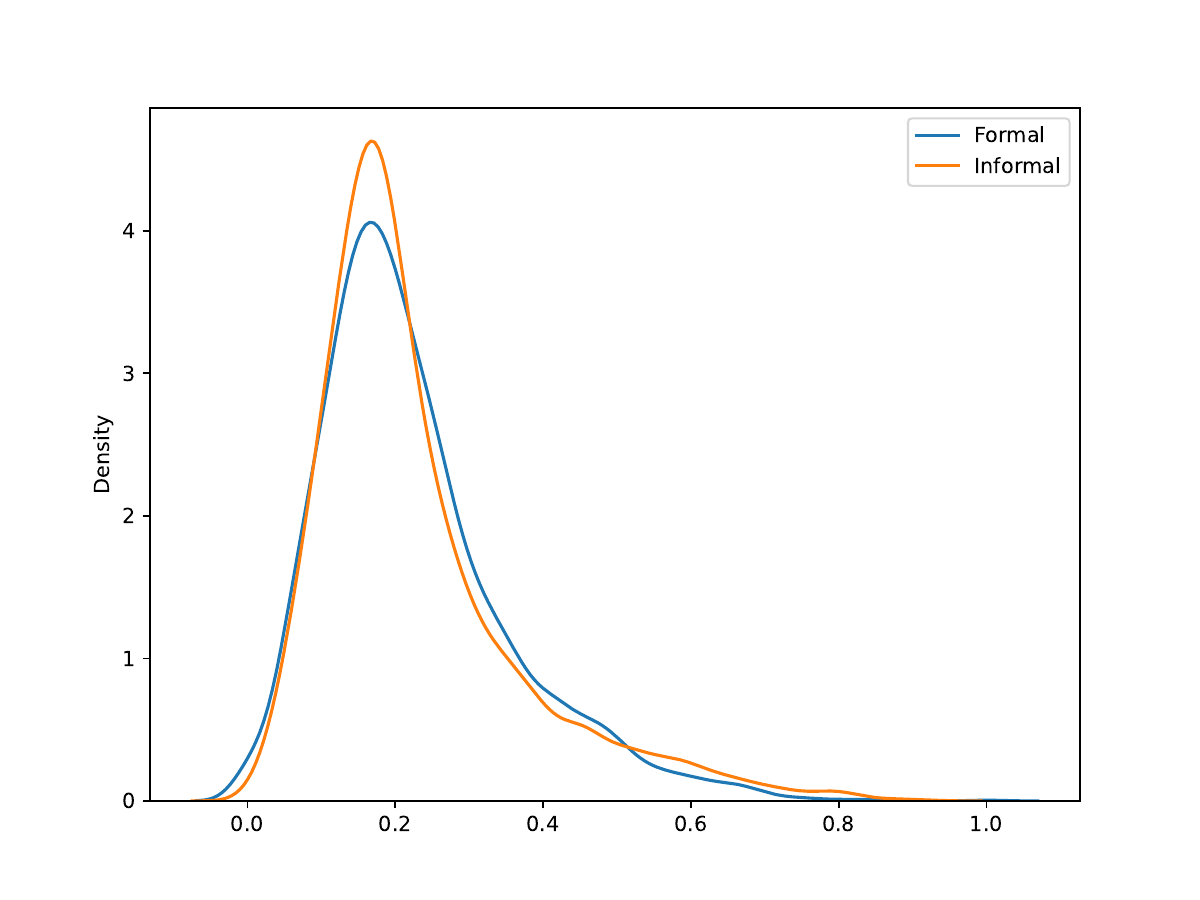} % Fourth figure
        \\
        Environmentally sustainable activities, 2000 (by sector)
    \end{minipage}

    \caption{Density function of activity employment shares across Brazilian municipalities. \\
    {\footnotesize Source: elaborated by the authors using data from Febraban and the Brazilian NSO (IBGE).  Notes: blue we see the density function for 2000 while in red the density function for 2010, using Census data.}}
    \label{fig:density-taxonomy}
\end{figure}

The share of environmentally risky activities in municipality $r$ is at period $t$ in total employment is: \\ $\sum_k \sum_{a \in \mathcal{R}(k)} L_{r,a,k,t-1} / L_{r,t-1}$. We plot these shares in Panel A of Figure \ref{fig:map-taxonomy} in the previous section. We also plot the density functions over time and across formalization profiles in Figure \ref{fig:density-taxonomy}. Environmentally risky activities decreased over time and were more prevalent across informal workers.

The following sectors have 100 percent of their activities classified as environmentally risky – i.e., subject to environmental licensing: rubber extraction; coal; oil; metallic and non-metallic minerals; leather products; wood products; paper and cellulose products; chemical products; pharmaceutical products; cement; metallurgy products; motors; mining equipment; handguns and rifles; appliances; cars, trucks, and buses; shipyards and shipbuilding; natural gas production and distribution; fuel retail trade; air transportation; sewage.\footnote{The list above is non-exhaustive. The full set of CNAE 3-digit industry codes (``grupos'') classified with 100 percent of environmentally risky activities is: 011, 012, 013, 014, 016, 017, 021, 051, 100, 111, 131, 132, 141, 142, 151, 152, 153, 154, 155, 156, 157, 158, 159, 160, 171, 172, 173, 174, 175, 176, 177, 181, 191, 192, 193, 201, 202, 211, 212, 213, 214, 231, 233, 234, 241, 242, 243, 244, 245, 246, 247, 248, 251, 252, 262, 263, 264, 269, 273, 274, 275, 281, 282, 283, 284, 289, 291, 292, 293, 294, 295, 296, 297, 298, 301, 302, 311, 312, 313, 314, 315, 316, 319, 321, 322, 323, 331, 332, 333, 334, 335, 341, 342, 343, 344, 351, 371, 372, 402, 410, 451, 456, 505, 601, 603, 611, 621, 622, 632, 900.}

To account for environmentally sustainable activities, the taxonomy replicates the European Union’s taxonomy for sustainable activities for Brazil’s national classification of economic activities (CNAE). The EU Taxonomy serves as a classification framework designed to guide firms and investors in identifying economic activities that qualify as environmentally sustainable. These activities are defined as those that contribute significantly to at least one of the EU's climate and environmental objectives while ensuring that they do not cause material adverse impacts on any other environmental goals and adhere to minimum social and governance safeguards.

The EU’s climate and environmental objectives include net zero by 2050; renewable energy transition; carbon pricing; biodiversity protection; circular economy; zero pollution; sustainable agriculture and food systems; improving forests and land use; and improving water and marine protection. The taxonomy aims to facilitate efficient capital allocation toward sustainable investments by providing standardized criteria for evaluating the environmental performance of economic activities. Since the taxonomy is directly derived from guidelines provided by the European Union, we take it to be plausibly exogenous to the spatial distribution of activities in Brazil. 

We classify activity $a$ of a 3-digit industry $k$ (i.e. sector) as environmentally sustainable if it is classified as ``high contribution'' to sustainability according to the taxonomy. We therefore do not classify as environmentally sustainable those activities with a ``moderate contribution'' to sustainability or those with no value assigned (no contribution to sustainability). The set of environmentally sustainable activities for each sector $k$ is, then:

\begin{equation*}
    \mathcal{S}(k) = \{ a : \text{ activity } a \text{ of sector } k \text{ has a high contribution to sustainability }\}  
\end{equation*}

At the three-digit industry level, the following sectors have 100 percent of their activities classified as environmentally sustainable: horticulture; medical machinery manufacturing; testing machinery manufacturing; optical machinery manufacturing; water treatment and distribution; intracity rail transportation; merchant marine transportation; database and online content distribution; research and development in natural sciences; research and development in social sciences; building cleaning services; social security administration; health care; social services; trade union services; artistic performances and concerts; libraries, archives, and museums.\footnote{These correspond to the following CNAE 3-digit industry codes (“grupos”): 012, 016, 331, 332, 334, 410, 601, 611, 724, 731, 732, 747, 753, 851, 853, 912, 923, 925.}

The share of environmentally sustainable activities in municipality r is at period t in total employment is: \\ $\sum_k \sum_{a \in \mathcal{S}(k)} L_{r,a,k,t-1} / L_{r,t-1}$. We plot these shares in Panel B of Figure \ref{fig:map-taxonomy} in the previous section. We also plot the density functions over time and across formalization profiles in Figure \ref{fig:density-taxonomy}. Environmentally sustainable activities are stable over time and equally prevalent across formal and informal workers.

\section{Results}\label{sec:results}
\begin{figure}[htp]
    \centering
    \includegraphics[width=0.75\linewidth]{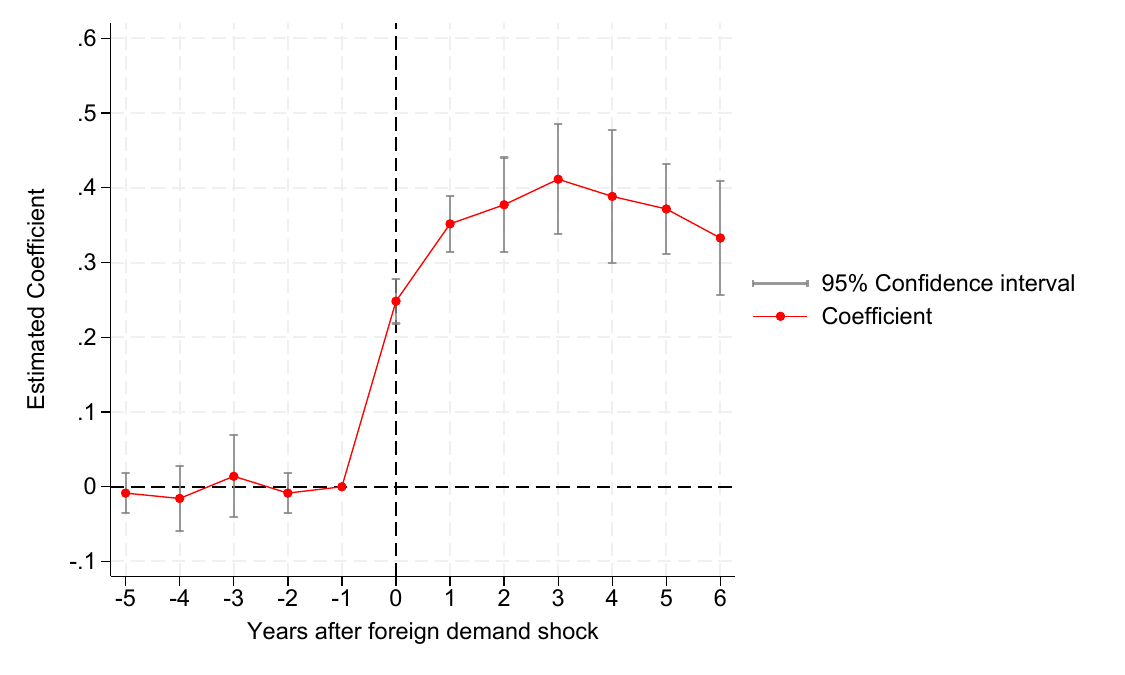}
    \caption{Elasticity of formal employment with respect to exports \\ {\footnotesize Source: elaborated by the authors using RAIS, UN COMTRADE, and MDIC data for the 2000-2020 period. Whiskers show 95\% confidence intervals based on robust standard errors clustered at the industry-group level.} }
    \label{fig:irf-employment}
\end{figure}

Our first result in Figure \ref{fig:irf-employment} shows that labor market adjustment is long-lasting after a positive demand shock in Brazil. The elasticity of formal employment to exports increases, peaks at around 0.4, and is still greater than 0.3 six years after the shock. We observe that in the five years leading up to the shock, there are no sizable differences across regions, suggesting that the growth of formal employment was similar in regions more and less exposed to the export shock. This absence of differential pre-trends is reassuring, suggesting that the shift-share instrumentation is likely valid, and the effects well identified. Regressing pre-shock outcomes on the instrument serve as balance tests that are recommended as best practice in the shift-share literature  \citep{borusyak_practical_2025, goldsmith-pinkham_bartik_2020}\footnote{As stated by \cite{borusyak_practical_2025}: ``Checking balance of the instrument at the unit level is relatively standard. For instance, a typical
pre-trend test involves regressing the lagged outcome on $z_i$ while including the controls picked in advance.''}.

This result confirms the relative slow adjustment of labor markets in Brazil documented in previous studies that leverage increased import competition due to the product market liberalization of the 1990s (e.g., \citealp{dix-carneiro_trade_2017}; \citealp{felix_trade_2022}). In that event, a negative shock induced regions more exposed to shocks to see a relative contraction in formal employment. In our setting, intuitively, we find the flip side of the coin: a positive demand shock induces a relative expansion in formal employment. We therefore complement the existing literature by showing that protracted labor market adjustment happens after positive demand shocks and beyond the specific liberalization event of the 1990s.

Recall that in the standard model the labor force is constant, and all the adjustment happens through average wages, since there is no migration and no unemployment. Here, we see adjustment happening through a relative expansion of formal employment in regions more exposed to the shock. We also see wages increasing more in those regions, which is consistent with being exposed to a stronger labor demand shock. 

\begin{figure}[htp]
    \centering
    \includegraphics[width=0.75\linewidth]{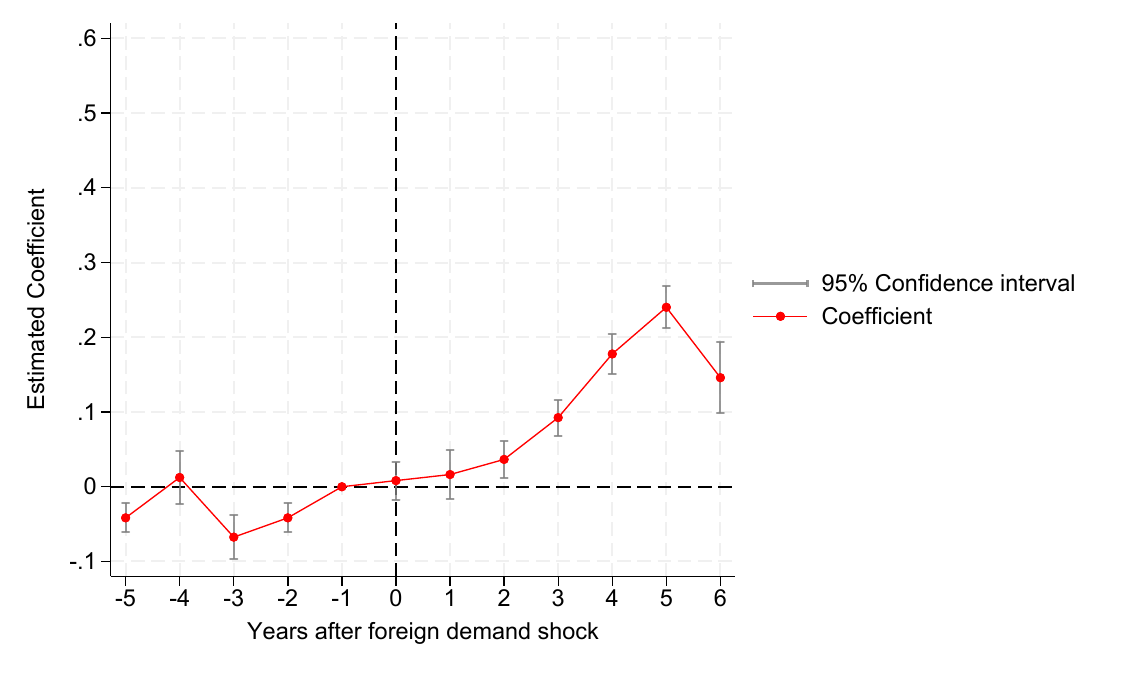}
    \caption{Elasticity of average wages with respect to exports \\ {\footnotesize Source: elaborated by the authors using RAIS, UN COMTRADE, and MDIC data for the 2000-2020 period. Notes: average wages were deflated using the yearly average deflator from IPCA, the Brazilian official CPI, at 2010 prices. Whiskers show 95\% confidence intervals based on robust standard errors clustered at the industry-group level.} }
    \label{fig:irf-wages}
\end{figure}

Consistent with labor market adjustment across regions, as the formal employment elasticity decreases, the elasticity of wages increases. As shown by Figure \ref{fig:irf-wages}, the effects are larger six years after the shock than in the very first period, suggesting that wages are likely sticky in the short run and need some time to adjust. One year after the foreign demand rise, average wages go up by 0.05 percent, on average, for every 1-percent exogenous increase in exports. Five years later, the estimated coefficient almost doubles, reaching approximately 0.2 percent.

\begin{figure}[htp]
    \centering
    \includegraphics[width=0.75\linewidth]{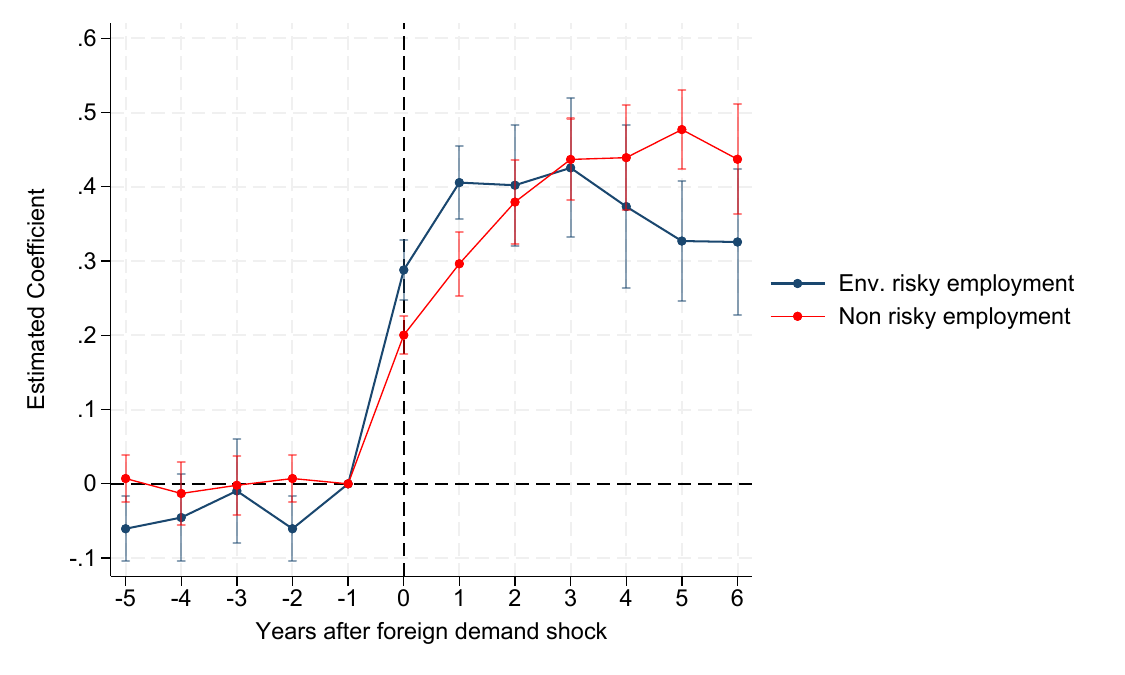}
    \caption{Elasticity of formal employment with respect to exports, separately for environmentally risky and non-risky activities \\ {\footnotesize Source: elaborated by the authors using RAIS, UN COMTRADE, and MDIC data for the 2000-2020 period. Whiskers show 95\% confidence intervals based on robust standard errors clustered at the industry-group level.} }
    \label{fig:irf-risky-emp}
\end{figure}

\paragraph{Exports, employment, and the environment} Over the short run, employment in activities subject to environmental licensing responds to export shocks more intensely than those not subject to same procedures (Figure \ref{fig:irf-risky-emp}). One year after the shock, the elasticity for environmentally risky employment is about 0.4, but the effect dissipates over time. This aligns with the overall perception that exports induce the expansion of ``dirty'' sectors in Brazil. However, over time the elasticity for non-environmentally risky employment peaks at 0.48, remaining higher than the elasticity of environmentally risky employment over the medium term. 

\begin{figure}[htp]
    \centering
    \includegraphics[width=0.75\linewidth]{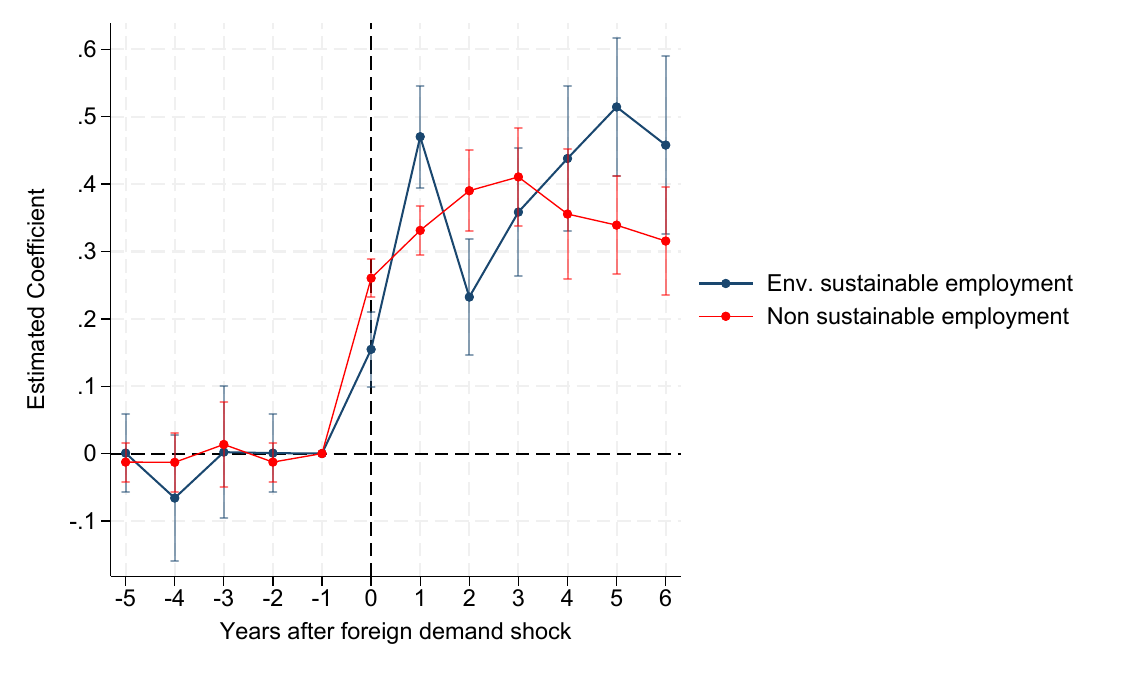}
    \caption{Elasticity of formal employment with respect to exports, separately for environmentally sustainable and non-sustainable activities \\ {\footnotesize Source: elaborated by the authors using RAIS, UN COMTRADE, and MDIC data for the 2000-2020 period. Whiskers show 95\% confidence intervals based on robust standard errors clustered at the industry-group level.} }
    \label{fig:irf-sust-emp}
\end{figure}

Environmentally sustainable employment elasticities show the reverse pattern. In regions more exposed to export shocks, sustainable employment responds (relatively) slower than non- sustainable employment over the short term but the effect over the long term seems to be more persistent. The peak elasticity of environmentally sustainable employment reaches 0.5, while that of activities not considered environmentally sustainable is 0.4.  The results are depicted in Figure \ref{fig:irf-sust-emp}.

Combined, these results suggest a nuanced picture regarding the impact of exports on the relative expansion of different sectors in Brazil. Immediately after the foreign demand shock, sectors not flagged as environmentally sustainable expand relatively faster. However, these are outpaced over time by the relative expansion of environmentally sustainable sectors. 

\paragraph{Exploring other heterogeneities} While the previous literature has shown that increased demand from China has impacted labor markets in Brazil, such as formalization rates and boost in employment \citep{costa_winners_2016}, we know little about the sectoral composition of this channel within formal employment. To shed some light on the mechanisms at play, we decompose the employment effect across different groups.

\textit{Skilled and non-skilled workers}. The differential effects are larger and more persistent for high- relative to low-skilled workers. On impact, formal employment of all skill levels located in regions more exposed to foreign demand shocks experience faster growth, with short-run elasticities around 0.25. Over the medium run, though, the elasticities are different. Six years after the foreign demand shock, formal employment elasticity of high skilled workers is 0.5, while that of low skilled workers converges to 0.3 (Appendix Figure \ref{fig:irf-employment-college}).

\textit{Male and female employment}. Over the short run, female and male formal employment respond in a similar way, with elasticities around 0.25. The relative response is higher for male workers over the medium run with male employment and female employment elasticities being around 0.4 and 0.3, respectively, six years after the shock (Appendix Figure \ref{fig:irf-employment-male}). 

\paragraph{Robustness} To ascertain the robustness of our baseline results, we use a different instrumental variable. Rather than using global exports as a proxy for foreign demand shocks, we calculate the industry-specific export-weighted exposure to demand growth in foreign destinations. We define our alternative instrument to be:

\begin{equation}\label{eq: alternative-instrument}
    \bar{\bar{x}}_{r,t} \equiv \sum_{k} \frac{L_{r,k,t-1}}{L_{r,t-1}} \sum_{d \notin \mathcal{B}} \tilde{\lambda}_{rd,k,t-1} \Delta y_{d,t}
\end{equation}

\noindent $\tilde{\lambda}_{rd,k,t-1} \equiv \lambda_{rd,k,t-1}/(\sum_{d' \notin \mathcal{B}} \lambda_{rd',k,t-1})$ is the export share of destination $d$ in industry $k$ at region $r$; $y_{d,t}$ is the log of dollar GDP in destination country $d$. Both the baseline and alternative instrument rely on the fact that each region in Brazil is small relative to the world, and global demand shocks are unlikely to be correlated with the distribution of domestic unobserved factors that differentially drive changes in local labor markets.

Both instrumental variables largely capture the same patterns, being distributed around an average of slightly positive growth rates and being strongly correlated. Appendix Figure \ref{fig:first-stage-appendix} shows the first-stage regression between the instrument and the instrumented variable, showing that the alternative instrument is relevant. Appendix Figure \ref{fig:dist-exposure} shows the instrument constructed from nominal dollar GDP growth rates has a longer left tail, possibly due to recessions in specific countries that do not translate into global export markets. Appendix Figure \ref{fig:correlation} shows that this alternative instrument is very correlated with the baseline instrument.

We chose the global exports-based instrument as our baseline because it is more strongly correlated with regional exports, especially for negative shocks. Appendix Figures \ref{fig:alternative-iv-exports} and \ref{fig:alternative-iv-wages} show the horizon specific elasticities of formal employment and wages with respect to exports using the alternative instrument. The results are qualitatively unchanged by using the alternative instrument.

\paragraph{Long-run effects and the informal labor market} As we mentioned in our description of the data, one of the limitations of the employee-employer matched dataset is that it excludes the informal sector – about half of the workforce in Brazil \citep{ulyssea_firms_2018}. In developing countries, this problem tends to be larger, as the informal labor market accounts for 20-80\% of workers \citep{ulyssea_informality_2020}. Therefore, we cannot speak about the heterogenous effects on formal and informal workers using these data alone. We expand on this topic below using Census data.

The census target sample is the universe of persons in Brazil and its window interval is approximately every ten years. It collects geographic, demographic, and labor market information (among others). To complement the analysis, we use two waves of the Census: 2000 and 2010. Those waves coincide with the large increase in aggregate exports in Brazil. After aggregating data at the municipality level, we run analogous models to equation \ref{eq:main-model}, using long differences. Our instrument is:

\begin{equation}\label{eq: gdp-instrument}
    \Delta \bar{x}_{r,2010} \equiv \sum_k \frac{L_{r,k,2000}}{L_{r,2000}} (\tilde{x}^w_{k,2010} - \tilde{x}^w_{k,2000})
\end{equation}

\noindent where $L_{r,k,2000}$ denotes total employment in industry $k$ at region $r$ in 2000; $L_{r,2000} = \sum_k L_{r,k,2000}$; and $\tilde{x}^w_{k,t}$ is the log of world exports (other than Brazil’s) in industry $k$ at period $t$. We then follow a similar 2SLS approach, with the first stage being:

\begin{equation*}
    \tilde{x}_{r,2010} - \tilde{x}_{r,2000} = \alpha + \beta \Delta \bar{x}_{r,2010} + \textbf{Z}_{r,2000}' \Gamma + \varepsilon_{r,2010}
\end{equation*}

\noindent where $\tilde{x}_{r,2010} - \tilde{x}_{r,2000}$ is the cumulative growth in exports of region r over that decade. The second stage:

\begin{equation}\label{eq:long-model}
    o_{r,2010} - o_{r,2000} = \alpha_{2010} + \beta_{2010} \Delta \hat{x}_{r,2010} + \textbf{Z}_{r,2000}' \Gamma_{2010} + \varepsilon_{r,2010}
    \end{equation}

where $ \Delta \hat{x}_{r,2010}$ are the predicted values of the first stage regression. Note that now we are using variation only along a cross-section of municipalities rather than exploiting the panel dimension as we were able to in the previous exercise. Despite this fact, the instrument is still quite relevant, with f-statistics above 30, suggesting that growth in foreign demand is a good predictor of growth in exports over longer time horizons.

In our preferred specification (Table~\ref{tab:lrelas}), the elasticity of formal employment to exports over this long horizon is 0.052, being statistically significant at the 1\% confidence level. This is consistent with the declining trend in elasticities as the horizon increases.\footnote{Note that there is a difference in interpretation in the long-differences specification compared to the dynamic version we presented in the previous section. While the latter shows the cumulative response of the outcome variable with respect to a single year's growth in exports, the former shows the cumulative response of the outcome variable with respect to the cumulative growth in exports. In other words, the marginal effect of the initial growth (say, between 2000 and 2001) is likely to be even smaller.} The response of the average wage of formal employees is also positive and statistically significant, albeit small, at 0.025. Taking full advantage of Census data, we also calculate the elasticity of informal employment with respect to exports. Notably, the elasticity is negative and statistically significant, meaning that regions more exposed to exports experience slower growth rates in informal employment relative to those least exposed to exports. While this might seem surprising, it is fully consistent with previous findings of the literature that documents labor market dynamics in Brazil.

For instance, \cite{dix-carneiro_trade_2017} find that informal employment grows at a relatively faster pace in regions more exposed to an import competition shock through product market liberalization in the 1990s. In a richer framework with informal and formal sectors, \cite{dix-carneiro_trade_2021} predict that the formal sector expands with higher foreign demand and that informal and formal sectors act as substitutes. If there are costs to formalization and imperfect enforcement, more productive firms sort into the formal sector (and vice versa), with the productivity threshold for formalization decreasing whenever the effective foreign market expands. In that sense, formal and informal employment seem to work as substitutes in Brazil, with the latter contracting and expanding in response to shocks to the former. The reason is that exporting firms are more likely to be in the formal sector and workers are unlikely to be very mobile in response to shocks, such that the adjustment happens primarily within the same region and across sectors.

\begin{table}[htp]
\centering
\begin{tabular}{lcccccc}
\multicolumn{7}{c}{\textbf{Long run elasticities with respect to exports}} \\\\ \hline
Sector & \multicolumn{2}{c}{Formal} & \multicolumn{2}{c}{Informal} & \multicolumn{2}{c}{Total} \\\\
\textbf{Dependent variables:} & (1) & (2) & (3) & (4) & (5) & (6) \\\\ \hline
Employment &  0.052 &  0.116 & -0.120 & -0.927 & -0.072 & -0.712 \\\\
          & ( 0.020) & ( 0.259) & ( 0.027) & ( 1.145) & ( 0.020) & ( 1.056) \\\\
First stage f-stat &  32.03 &   0.33 &  34.25 &   0.66 &  33.45 &   0.47 \\\\
N &      1278 &      1270 &      1278 &      1270 &      1278 &      1270 \\\\ \hline
Environmentally risky employment &  0.072 &  0.074 & -0.152 & -1.117 & -0.103 & -1.050 \\\\
          & ( 0.027) & ( 0.280) & ( 0.034) & ( 1.344) & ( 0.028) & ( 1.575) \\\\
First stage f-stat &  31.32 &   0.29 &  34.12 &   0.69 &  33.19 &   0.46 \\\\
N &      1278 &      1270 &      1278 &      1270 &      1278 &      1270 \\\\ \hline
Environmentally sustainable employment &  0.171 &  0.067 &  0.040 &  0.030 &  0.083 & -0.164 \\\\
          & ( 0.041) & ( 0.304) & ( 0.030) & ( 0.234) & ( 0.028) & ( 0.357) \\\\
First stage f-stat &  32.82 &   0.35 &  34.88 &   0.56 &  34.38 &   0.40 \\\\
N &      1278 &      1270 &      1278 &      1270 &      1278 &      1270 \\\\ \hline
Real wages &  0.025 &  0.021 &  0.049 &  0.256 &  0.044 &  0.303 \\\\
          & ( 0.012) & ( 0.080) & ( 0.019) & ( 0.310) & ( 0.013) & ( 0.375) \\\\
First stage f-stat &  32.30 &   0.67 &  33.40 &   0.77 &  32.68 &   0.66 \\\\
N &      1278 &      1270 &      1278 &      1270 &      1278 &      1270 \\\\ \hline
State Fixed Effects & Y & N & Y & N & Y & N \\\\ \hline
\end{tabular}

\caption{Impacts on employability and real average wages, considering formal and informal workers separately \\ {\footnotesize Source: elaborated by the authors using the Brazilian Population Census data for 2000 and 2010. Notes: formal workers are those with a formal job contract or who contribute to social security. Robust standard errors clustered at the industry-group level in parenthesis.}}
\label{tab:lrelas}
\end{table}

Consistent with the results on the previous sections, the elasticity of formal environmentally risky employment drops substantially as the horizon increases to 0.07 after 10 years. However, environmentally sustainable employment elasticities are much more persistent over time, at 0.17. Exports growth decrease the relative growth of informal environmentally risky employment with an elasticity of -0.15. However, it has no effect over informal environmentally sustainable employment. All in all, our results suggest that the widespread perception that a growth in the exporting sectors in Brazil would lead to the expansion of environmentally harmful sectors does not hold over the long run.

\section{Conclusions}\label{sec:conclusion}
This paper examines the impacts of foreign demand shocks on local labor markets in Brazil over a 20-year period, focusing on the role of export growth in driving labor market outcomes. Between 2000 and 2020, Brazil experienced a tripling of exports, predominantly in the agriculture, oil, and mining sectors. The exposure to these foreign demand shocks varied significantly across the country, with some regions being more exposed to export-driven growth than others. By leveraging this regional variation, we show that municipalities that face increased exports experience faster growth in formal employment. These elasticities are 0.25 on impact, peak at 0.4 and remain positive and significant even 10 years after the shock, pointing to a long and protracted labor market adjustment.

Despite these successes in formal job creation, the persistence of labor informality remains a challenge for Brazil. Many workers, particularly those in lower-skilled roles, continue to find employment in the informal sector, which offers fewer protections and lower wages. Our findings suggest that regions with higher exposure to export shocks are less likely to experience increases in informal employment, suggesting that trade can help formalize the labor market in the short term. This is consistent with broader trends observed in trade liberalization research, where informal sectors act as a buffer for displaced workers.  

The benefits of export growth also appear to be distributed relatively evenly across genders, with formal employment elasticities being higher for males but wage elasticities being higher for females. Effects are more pronounced for highly skilled workers, who benefit from longer-lasting wage and employment gains compared to lower-skilled workers. The paper's findings align with other empirical research that highlights how export-driven sectors, particularly those with higher skill requirements, are more resilient to economic shocks and tend to provide better wages and job stability.

To investigate the relationship between exports and the environment, we combine labor market and trade data with a granular taxonomy of economic activities to estimate employment elasticities with respect to exports in sectors with different environmental characteristics. Using census data, we document that environmentally risky activities have a larger share of employment than environmentally sustainable ones. The relationship between those activities and exports is nuanced. In the short run, the employment elasticity of environmentally risky employment is higher than that of environmentally sustainable employment. However, the pattern flips over time. Even ten years after the initial shock, the elasticity of formal environmentally sustainable employment stands at 0.17.

While the short-term gains from export growth are substantial, the benefits tend to dissipate over time, particularly in sectors with higher environmental risks. Industries such as agriculture, oil, and mining initially drive formal job creation in response to export shocks, but this effect diminishes within four years. This aligns with other studies suggesting that trade shocks often result in temporary boosts to employment, followed by market adjustments that neutralize these gains over time.

Considering these findings, it is crucial to consider how export growth interacts with environmental sustainability. While sectors with higher environmental risks, such as mining and agriculture, are key drivers of export-led job creation in Brazil in the short run, these effects dissipate four years after the shock. However, the long-term viability of these industries may be challenged by global efforts to reduce carbon emissions and transition toward greener economies. Brazil, as one of the world's largest exporters of commodities, has a unique opportunity to leverage its trade growth while shifting toward sustainable practices. Policies that promote green jobs, such as investments in renewable energy, sustainable agriculture, and eco-friendly manufacturing, could help balance economic growth with environmental goals.

To ensure that the benefits of trade and structural transformation are both inclusive and sustainable, Brazil needs to implement targeted policies that not only foster formal employment but also support the green transition. By promoting labor formalization, improving job quality, and incentivizing greener industries, Brazil can better align its trade policies with its long-term development goals. These efforts will be essential in ensuring that the economic gains from export growth are widely distributed across the population while positioning Brazil as a leader in the global movement toward a greener and more sustainable economy.

\clearpage
\bibliographystyle{bib/aeaown}
\bibliography{bib/references}

\clearpage
\appendix

\section{Additional figures}\label{sec:figures}
%\clearpage
\setcounter{figure}{0}
\renewcommand{\thefigure}{A.\arabic{figure}}

\begin{figure}[ht]
    \centering
    \includegraphics[width=0.75\linewidth]{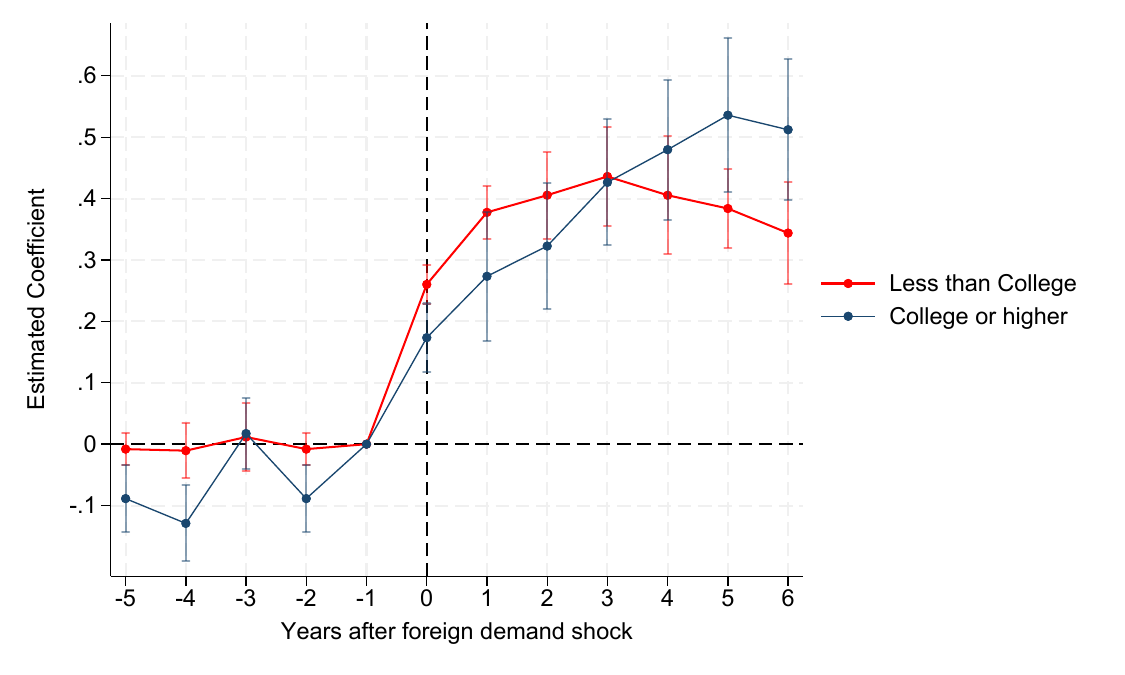}
    \caption{Elasticity of formal employment with respect to exports, by schooling level of workers \\ {\footnotesize Source: elaborated by the authors using RAIS, UN COMTRADE, and MDIC data for the 2000-2020 period. } }
    \label{fig:irf-employment-college}
\end{figure}

\clearpage

\begin{figure}[htp]
    \centering
    \includegraphics[width=0.75\linewidth]{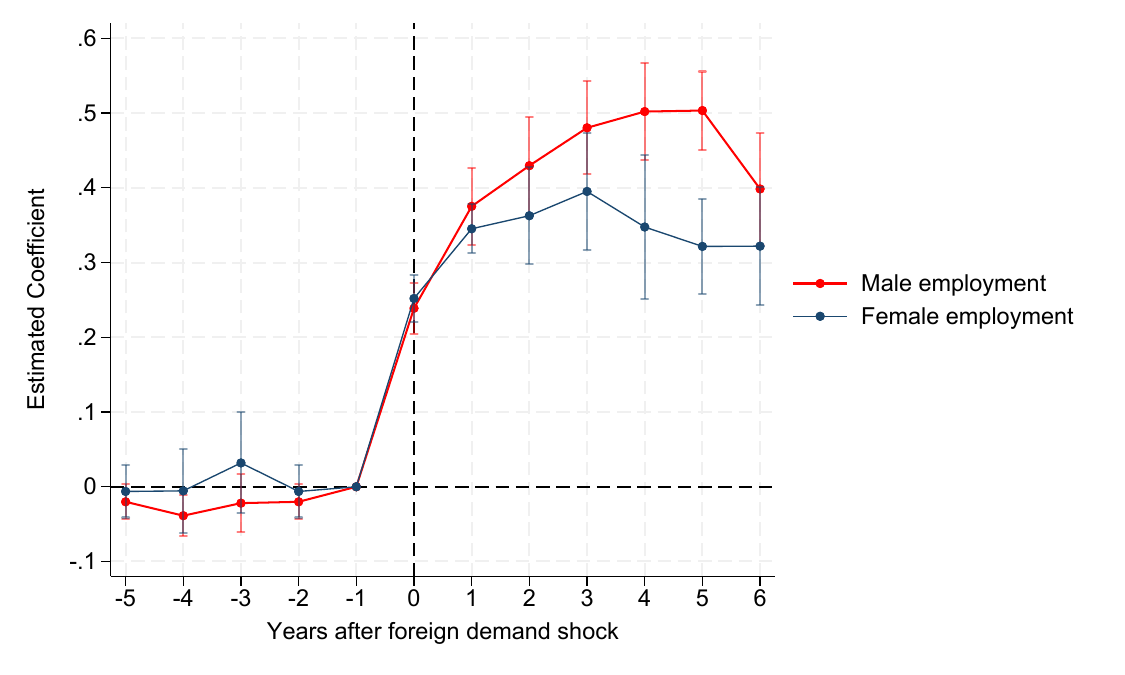}
    \caption{Elasticity of formal employment with respect to exports, by gender of workers \\ {\footnotesize Source: elaborated by the authors using RAIS, UN COMTRADE, and MDIC data for the 2000-2020 period. } }
    \label{fig:irf-employment-male}
\end{figure}

\clearpage

\begin{figure}[htp]
    \centering
    \includegraphics[width=0.75\linewidth]{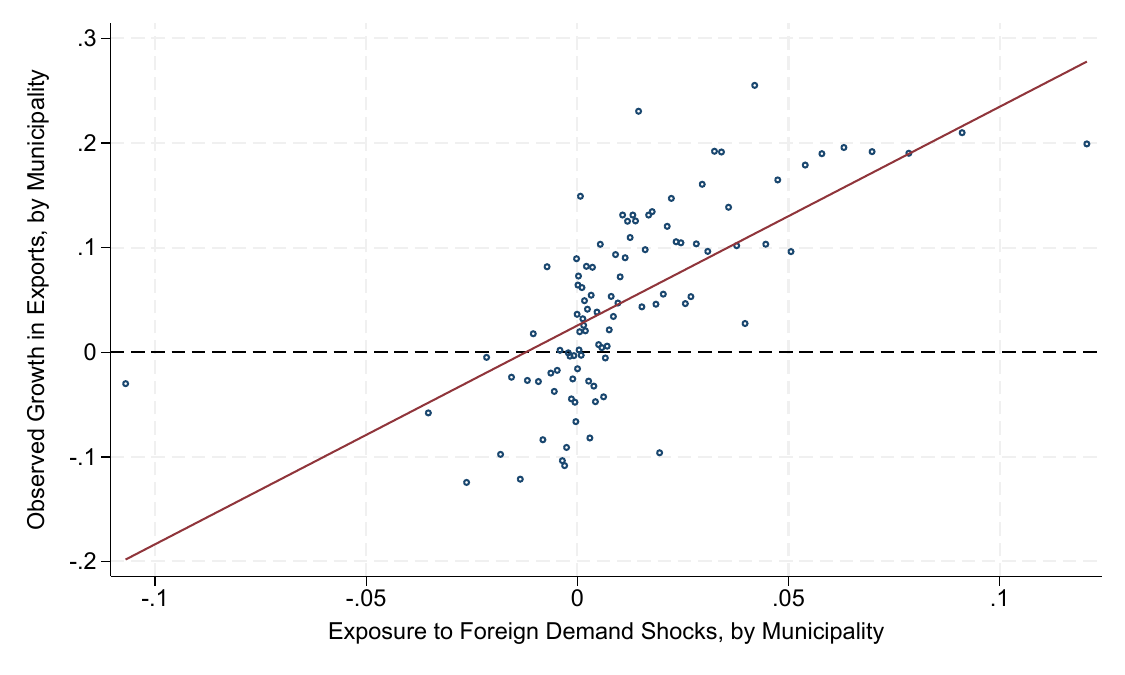}
    \caption{First stage of two-stage least squares, using the alternative instrument \\ {\footnotesize Source: this binscatter reproduces the slope of regressing the observed growth in exports on the alternative instrument. The underlying regression has N=35,729, Beta = 2.09 and t-stat=10.29.} }
    \label{fig:first-stage-appendix}
\end{figure}

\clearpage

\begin{figure}[htp]
    \centering
    \includegraphics[width=0.75\linewidth]{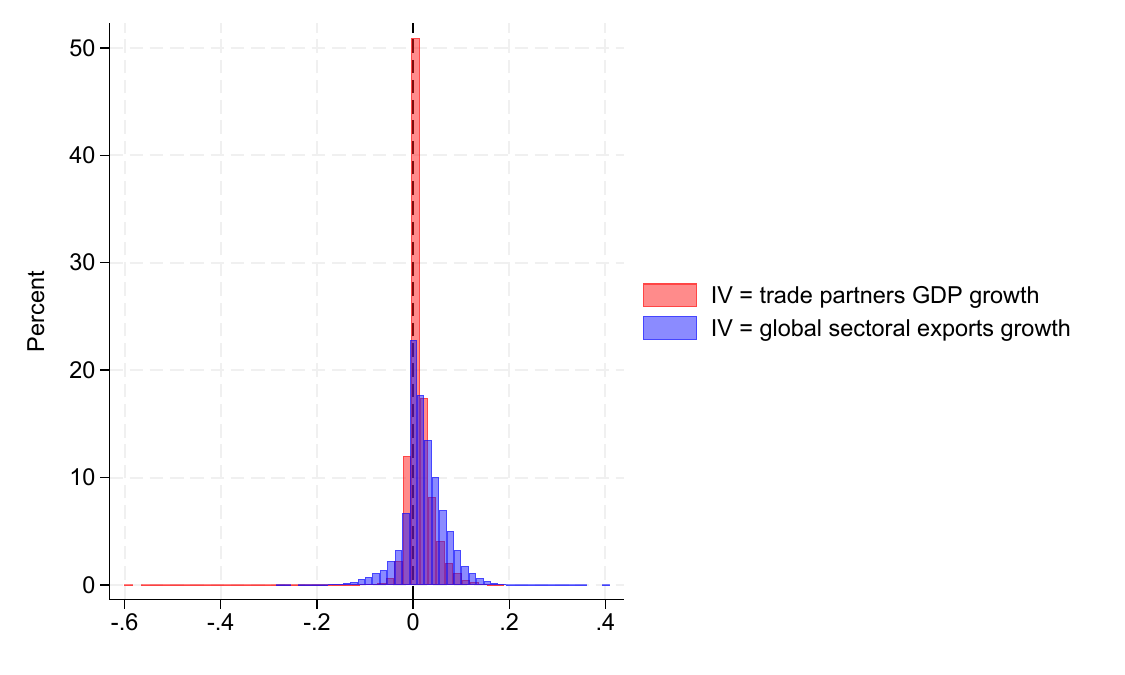}
    \caption{Distribution of exposure to foreign demand shocks by municipality \\ {\footnotesize Source: elaborated by the authors using RAIS, UN COMTRADE, and MDIC data for the 2000-2020 period. } }
    \label{fig:dist-exposure}
\end{figure}

\clearpage

\begin{figure}[htp]
    \centering
    \includegraphics[width=0.75\linewidth]{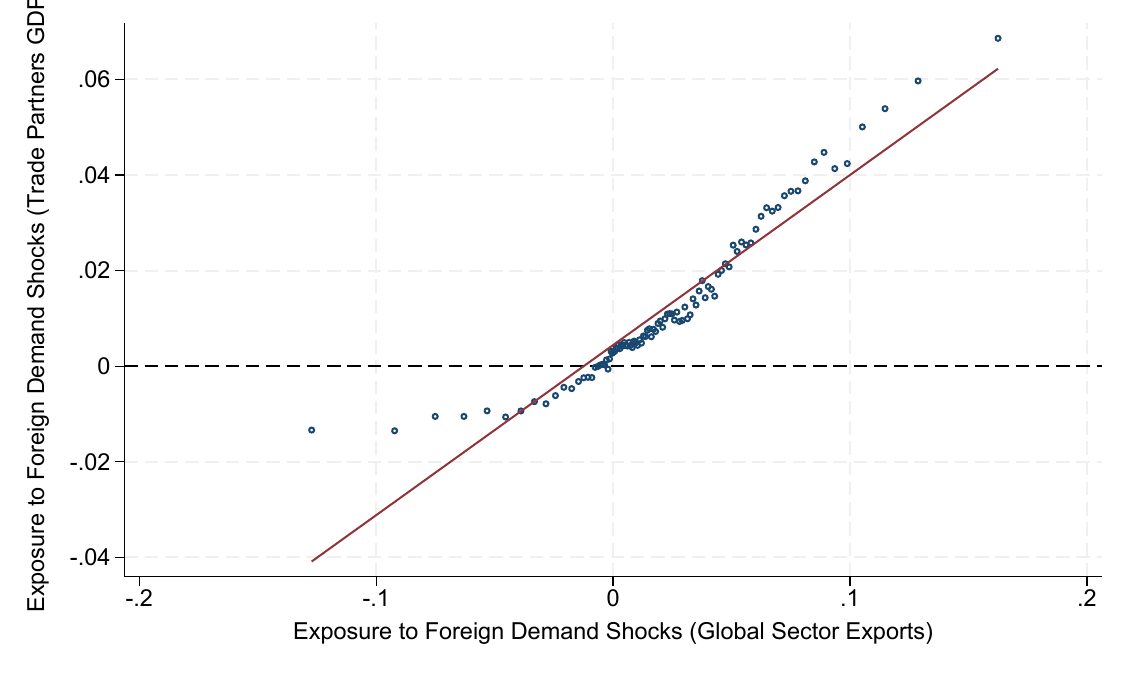}
    \caption{Correlation between two instrumental variables \\ {\footnotesize Source: elaborated by the authors using RAIS, UN COMTRADE, and MDIC data for the 2000-2020 period. } }
    \label{fig:correlation}
\end{figure}

\clearpage

\begin{figure}[htp]
    \centering
    \includegraphics[width=0.75\linewidth]{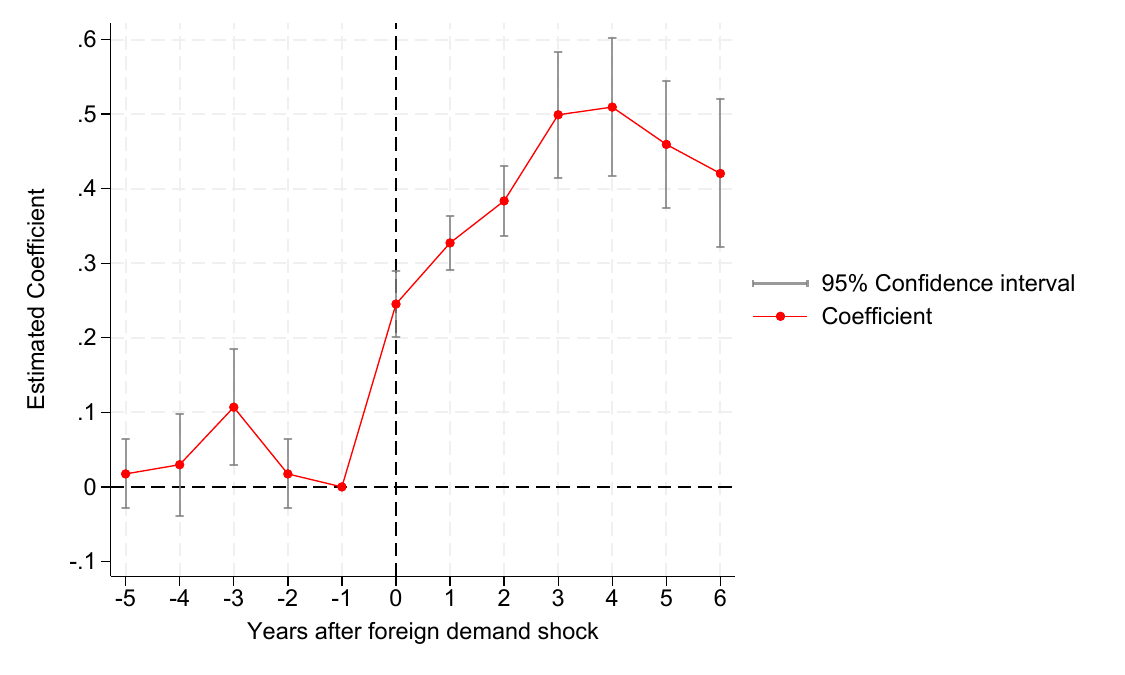}
    \caption{Alternative instrument: elasticity of employment with respect to exports \\ {\footnotesize Source: elaborated by the authors using RAIS, UN COMTRADE, and MDIC data for the 2000-2020 period. } }
    \label{fig:alternative-iv-exports}
\end{figure}

\clearpage

\begin{figure}[htp]
    \centering
    \includegraphics[width=0.75\linewidth]{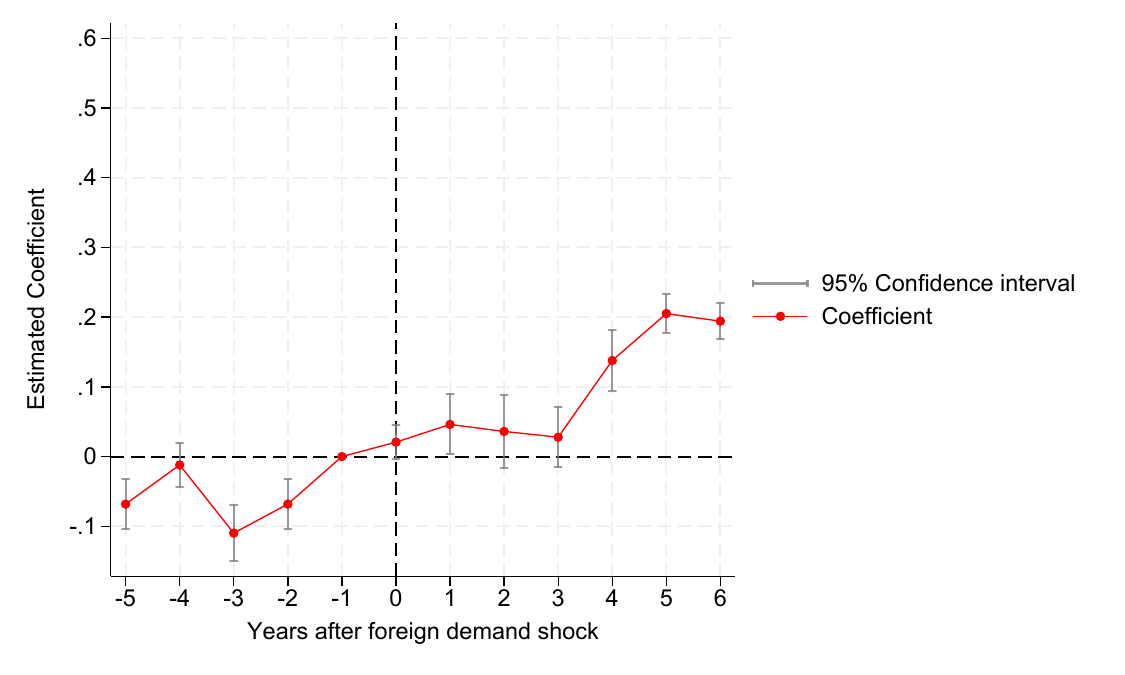}
    \caption{Alternative instrument: elasticity of wages with respect to exports \\ {\footnotesize Source: elaborated by the authors using RAIS, UN COMTRADE, and MDIC data for the 2000-2020 period. } }
    \label{fig:alternative-iv-wages}
\end{figure}

\clearpage

\section{Mapping the taxonomy of economic activities to labor data}\label{sec:app-febraban}

\paragraph{Description of the taxonomy of economic activities}	
In 2020, the Brazilian Banks Federation (FEBRABAN) introduced a taxonomy with two variables to classify economic activities in Brazil based on environmental considerations. This initiative aimed to determine whether the credit provided by banks was directed more towards environmentally risky sectors or towards sustainable ones. The taxonomy was developed through discussions of the Working Group on Climate and Green Economy, which included fifteen banks and was coordinated by FEBRABAN. It is based on the latest official classification of economic activities in Brazil (CNAE 2.0) at its most detailed level (7-digit industry level), classifying 1,331 economic activities according to the two variables.

The first variable, environmentally sustainable, identifies economic activities considered by the United Nations Environment Programme as welfare-enhancing and environmentally friendly due to low carbon emissions and efficient use of environmental resources. FEBRABAN drew inspiration from existing categorizations for sustainable activities, including the European Union’s taxonomy, the Climate Bonds Initiative, and the Green Bond Principles from the International Capital Markets Association. This variable can take on two values: “high contribution” or “moderate contribution.” If an economic activity does not significantly contribute to the environment’s sustainability, this variable has a missing value. According to FEBRABAN’s classification, 168 activities had a high contribution, 112 activities had moderate contribution and 1,051 activities had a missing value.

The second variable, environmentally risky, is based on a 1997 normative decree from the Brazilian government, which mandates environmental licensing for economic activities with high potential environmental impact. This decree does not list the CNAE codes of the activities requiring licensing, so FEBRABAN interpreted the law to create this variable. It can only take on one value: “high exposure.” If an economic activity is not considered to have a high potential environmental impact, this variable is left blank. FEBRABAN classified 552 activities as being high exposure.

\paragraph{Mapping the taxonomy of sectors to RAIS data}
The national classification of economic activities (CNAE) in Brazil was officially launched in 1995 and its most detailed level was at the 5-digit industry level. It underwent a significant revision in 2007, when it became CNAE 2.0, but before that the Brazilian national statistical office (IBGE) launched another categorization (CNAE 1.0). To ensure comparability across years, the RAIS data for the 2000-2020 period is based on the first version of CNAE (CNAE 95) at the 5-digit industry level. Mapping FEBRABAN’s taxonomy - based on CNAE 2.0 at the 7-digit industry level - to RAIS data requires certain assumptions.
IBGE provides an official concordance from CNAE 2.0 to CNAE 1.0 at the 5-digit level on both sides and from CNAE 1.0 to CNAE 95 (also at the 5-digit level on both sides), but there is no direct mapping from CNAE 2.0 to CNAE 95. Therefore, a multi-stage procedure was necessary. First, each of FEBRABAN’s variables of interest (which followed the CNAE 2.0 categorization at the 7-digit level) was transformed into a 5-digit level. If at least one 7-digit activity was considered “high exposure” in the environmental risk variable, the corresponding 5-digit industry level code was classified as “high exposure,” regardless of the number of 7-digit codes associated with it. A similar procedure was applied to the environmentally sustainable variable: if at least one 7-digit activity was considered ‘high contribution’, the corresponding 5-digit industry level code was classified as environmentally sustainable. Once both variables from FEBRABAN’s taxonomy were mapped to the CNAE 1.0 categorization at the 5-digit level, the next step was to use the concordance from CNAE 1.0 to CNAE 95. Out of a total of 560 5-digit industries classified under CNAE 95, 322 industries were identified as environmentally risky and 87 as environmentally sustainable.

\paragraph{Mapping the taxonomy of sectors to Census data}
Mapping FEBRABAN’s taxonomy to census data for the years of 2000 and 2010 involves additional steps beyond those described for RAIS data. These additional steps are related to the fact that the 2010 Population Census microdata uses a classification of economic activities named CNAE Domiciliar 2.0 that is different from the one used in the 2000 Population Census called CNAE Domiciliar 1.0. Both are distinct from CNAE 95 used in RAIS data and from FEBRABAN’s 7-digit industry classification.
To map FEBRABAN’s classification to census data, other concordances published by IBGE were used. Two additional concordances are available: one from CNAE Domiciliar 1.0 to CNAE 95 and another from CNAE Domiciliar 2.0 to CNAE Domiciliar 1.0. For the 2000 Population Census, the steps described for mapping to RAIS data were followed, along with using IBGE’s concordance from CNAE Domiciliar 1.0 to CNAE 95. For the 2010 Population Census, we followed the steps described for mapping to 2000 Census and additionally used the concordance from CNAE Domiciliar 2.0 to CNAE Domiciliar 1.0.

\end{document}